\documentclass[twoside]{article}
\usepackage{epsfig,fleqn,espcrc2}

\usepackage{graphicx}
\usepackage[figuresright]{rotating}

\newcommand{\AmS}{{\protect\the\textfont2
  A\kern-.1667em\lower.5ex\hbox{M}\kern-.125emS}}

\def\Dslash{{D \!\!\!\!/}}

\def\efb{f_B}
\def\efbs{f_{B_s}}

\def\bbbar{$B$-$\bar B$}
\def\fb{$f_B$}
\def\fbs{$f_{B_s}$}
\def\fd{$f_D$}
\def\fds{$f_{D_s}$}
\def\fbsofb{$f_{B_s}/f_B$}
\def\ehbbd{\hat B_{B_d}}
\def\ehbbs{\hat B_{B_s}}
\def\hbbd{$\hat B_{B_d}$}

\def\gtwid{\raise.3ex\hbox{$>$\kern-.75em\lower1ex\hbox{$\sim$}}}
\def\ltwid{\raise.3ex\hbox{$<$\kern-.75em\lower1ex\hbox{$\sim$}}}

\def\ie{{\it i.e.},\ }
\def\eg{{\it e.g.},\ }
\def\et{{\it et al.}}

\def\cM{{\cal M}}

\def\cO{{\cal O}}
\def\cP{{\cal P}}

\def\prl#1{Phys.\ Rev.\ Lett.\ {\bf #1}}
\def\prd#1{Phys.\ Rev.\ {\bf D#1}}
\def\plb#1{Phys.\ Lett.\ {\bf #1B}}
\def\npb#1{Nucl.\ Phys.\ {\bf B#1}}

\def\seillac{Nucl.\ Phys.\ {\bf B} (Proc.\ Suppl.) {\bf 4} (1988)}

\def\edinburgh{Nucl.\ Phys.\ {\bf B} (Proc.\ Suppl.) {\bf 63} (1998)}
\def\boulder{Nucl.\ Phys.\ {\bf B} (Proc.\ Suppl.) {\bf 73} (1999)}
\def\pisa{Nucl.\ Phys.\ {\bf B} (Proc.\ Suppl.) {\bf 83-84} (2000)}

\def\MeV{{\rm Me\!V}}
\def\GeV{{\rm Ge\!V}}

\title{Heavy quark physics on the lattice}

\author{C.\ Bernard\vskip 6pt%
{Department of Physics, Washington University,
        St.\ Louis, MO 63130}}
       
\begin{document}

\begin{abstract}
I review the current status of lattice calculations of the properties
of bound states containing one or more heavy quarks. 
Many of my remarks focus on
the heavy-light leptonic decay constants, such as $f_B$, for which 
the systematic errors have by now been quite well studied.
I also discuss $B$-parameters, semileptonic
form factors, and the heavy-light and heavy-heavy spectra.
Some of my ``world averages'' are: $f_B=200(30)\;\MeV$,
$f_B\sqrt{\hat B_{B_d}}= 230(40)\;\MeV$, $f_{B_s}/f_B=1.16(4)$ and
$f_{B_s}\sqrt{\hat B_{B_s}}/f_B\sqrt{\hat B_{B_d}}=1.16(5)$.
\vspace{-1pc}
\end{abstract}

\maketitle

\section{INTRODUCTION}
Many of the parameters of the Standard Model can be 
constrained by measurements of the properties of hadrons 
containing heavy quarks.  To take advantage of such experiments, 
however, one needs theoretical
determinations of the corresponding strong-interaction matrix
elements.  Lattice gauge theory provides, at least in principle,
a means of computing hadronic matrix elements with control over
all sources of systematic error. 
Here, I review the current status of these computations.

A review like this is necessarily somewhat idiosyncratic in the
topics it covers and the time spent on each.  
I devote a large
fraction of my time to $f_B$.  
Because lattice
data for leptonic decay constants is more extensive 
than for any other heavy quark  quantity,
I am able to discuss
in detail several vexing issues:
renormalon effects, systematic errors due to large lattice quark masses,
extrapolation to the continuum, and quenching errors.  
I then turn to the ``$B$-parameters,''
particularly $B_{B_d}$ and $B_{B_s}$, and to semileptonic form
factors for $B\to\pi$. Finally, I discuss spectra of heavy-light
and heavy-heavy hadrons.
I conclude
with a few remarks on the unitarity triangle.  

Heavy quark masses, which are included in Lubicz's talk at
this conference \cite{LUBICZ},  are examined here only for the
light they shed on renormalon effects. 
The heavy quark potential is discussed in detail by
Bali \cite{BALI}, and I omit it here.
Recent reviews of heavy quark physics
on the lattice by Hashimoto \cite{HASHIMOTO_LAT99} and Draper 
\cite{DRAPER_LAT98} complement the current treatment.

\section{LEPTONIC DECAY CONSTANTS}
\label{sec:fB}

\begin{figure}[htb]
\null
\vspace{.1truein}
\includegraphics[bb = 50  100 750 550,
width=3.2truein,height=1.9truein]{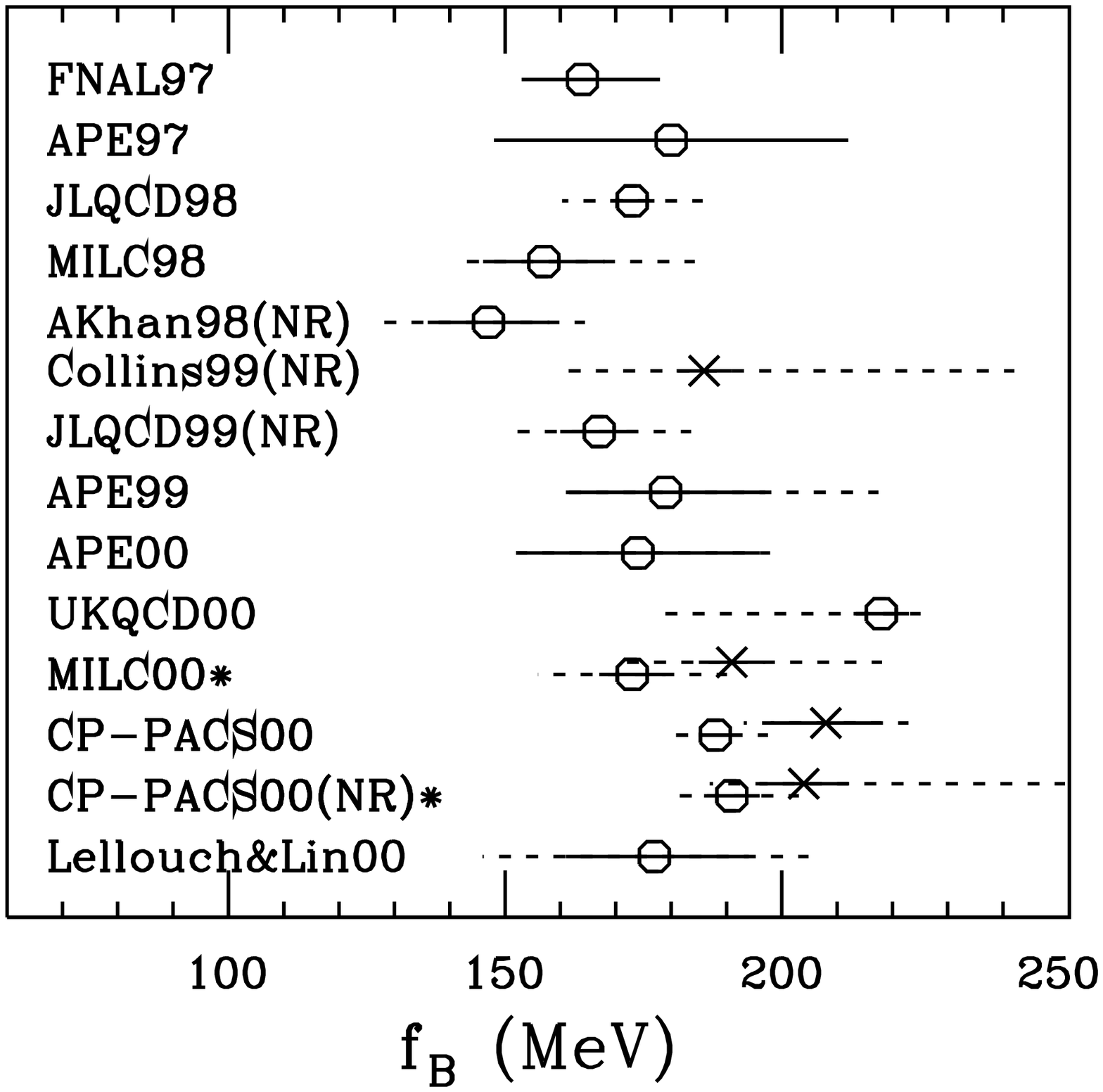}
\vskip -.1truein
\caption{Results for $f_B$ from various groups in the
quenched approximation (circles) and with $N_f=2$ (crosses).
Solid lines show statistical errors; dashed lines show systematic and
statistical errors, added in quadrature.
A * denotes a preliminary result. Calculations
using NRQCD are marked (NR); all other computations using relativistic
formalisms. For references see
Table \ref{tab:fb}.} 
\label{fig:compare_fb}
\vspace{-0.05truein}
\end{figure}

\begin{figure}[htb]
\null
\vspace{-0.25truein}
\includegraphics[bb = 100  600 4096 4196,
width=2.5truein,height=2.3truein]{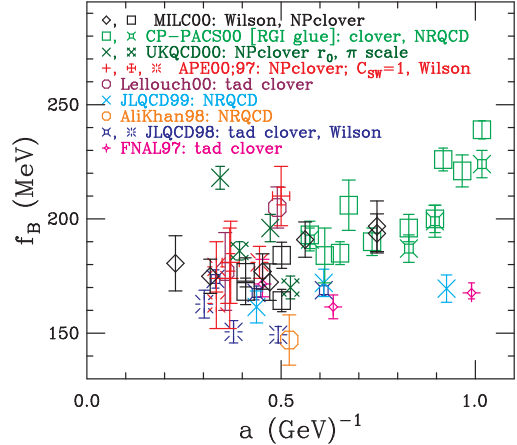}
\caption{Recent world data for $f_B$  in the quenched approximation
 {\it vs.}\ lattice spacing. Only statistical errors are shown. 
For references see
Table \ref{tab:fb}.}
\label{fig:fb_quenched_allgroups}
\vspace{-0.1truein}
\end{figure}

Table \ref{tab:fb} and Fig.~\ref{fig:compare_fb} show recent
results for $f_B$; while
Fig.~\ref{fig:fb_quenched_allgroups} plots the quenched $f_B$ data
as a function of lattice spacing $a$.
One notices immediately
that 
the UKQCD \cite{UKQCD_fB00,MAYNARD} and 
CP-PACS results \cite{CPPACS_fB00,CPPACS_NEW} (with heavy clover
quarks and with NRQCD) are somewhat
high compared to those of the other groups. Furthermore,
the errors given by CP-PACS are rather small.
Does this mean one should increase the central value and
lower the error significantly from the ``world
averages'' for quenched $f_B$ quoted recently by
and Hashimoto \cite{HASHIMOTO_LAT99} ($f_B=170(20)\;\MeV$) or
Draper \cite{DRAPER_LAT98} ($f_B=165(20)\;\MeV$)?

\begin{table*}[htb]
\begin{center}
\begin{tabular}{|c|c|c|}
\hline  
Quenched& $f_{B}$ (MeV)&
$f_{B_s}/f_{B}$ 
\\
\hline 
FNAL97 \cite{FNAL97}& $164(^{+14}_{-11})(8)$  & $1.13(^{+5}_{-4})$ 
\\
APE97 \cite{APE97}& $180(32)$  & $1.14(8)$ 
\\
JLQCD98 \cite{JLQCD98}& $173(4)(12)$  & $\simeq 1.15$ 
\\
MILC98 \cite{MILC_PRL}& $157(11)(^{+25}_{-9})(^{+23}_{-0})$  
& $1.11(2)(^{+4}_{-3})(3)  $
\\
Ali Khan98 \cite{ALIKHAN98}& $147(11)(^{+8}_{-12})(9)(6)$ & $1.20(4)(^{+4}_{-0})$
\\
JLQCD99 \cite{JLQCD99} & $167(7)(15)$ & $1.15(3)(1)(^{+3}_{-0})$
\\
APE99 \cite{APE99}& $179(18)(^{+34}_{-9})$ & $1.14(2)(1)$ 
 \\
APE00 \cite{APE_fB_BB00}& $174(22)(^{+7}_{-0})(^{+4}_{-0})$ & $1.17(4)(^{+0}_{-1})$ 
 \\
UKQCD00 \cite{UKQCD_fB00,MAYNARD}& $218(5)(^{+5}_{-41})$ & $1.11(1)(^{+5}_{-3})$
\\
MILC00$^*$ \cite{MILC_LAT00} &$173(6)(16)$  
& $1.16(1)(2) $
\\
CP-PACS00 \cite{CPPACS_fB00,CPPACS_NEW}& $188(3)(9)$ &
$1.148(8)(46)(^{+39}_{-0})$ 
\\  
CP-PACS00(NR)$^*$ \cite{CPPACS_fB00}& $191(5)(11)$ &
$1.150(9)(6)$ 
\\  
Lellouch\&Lin00 \cite{LELLOUCH00}& $177(17)(^{+22}_{-26})$ &
$1.15(2)(^{+3}_{-2})$  
 \\
\hline 
\hline 
$N_f=2$& $f_{B}$ (MeV)& $f_{B_s}/f_{B}$ 
\\
\hline 
Collins99(NR) \cite{COLLINS99}& $186(5)(19)(9)(13)(^{+50}_{-0})$ & $1.14(2)(^{+0}_{-2})$ 
 \\
MILC00$^*$ \cite{MILC_LAT00} & $191(6)(^{+24}_{-18})(^{+11}_{-0})$ &
$1.16(1)(2)(2)$
 \\
CP-PACS00   \cite{CPPACS_fB00,CPPACS_NEW}& $208(10)(11)$ &
$1.203(29)(43)(^{+38}_{-0})$ \\  
CP-PACS00(NR)$^*$ \cite{CPPACS_fB00}& $204(8)(15)(^{+44}_{-0})$ &
$1.179(18)(7)$ \\  
\hline 
\end{tabular} 
\caption[]{ 
$f_B$ and $f_{B_s}/f_B$ from various groups.   Computations
using NRQCD are explicitly 
noted with (NR); all others use relativistic formalisms.
A * denotes a preliminary result.
}
\label{tab:fb} 
\end{center}
\vskip -.15truein
\end{table*}

In the case of UKQCD, a major cause of the quite high result is the choice
of the $r_0$ parameter \cite{SOMMER93} from the static potential to 
set the scale.  Since $ar_0$ can be
extracted easily and precisely from lattice data, it provides
an excellent way to check the scaling of physical quantities
as $a$ is changed. However, I would argue that because $r_0$ is not
directly determined by experiment, but only through phenomenology,
$ar_0$ should not be used to set the absolute scale of $a$. Indeed,
Sommer \cite{SOMMER93} remarks that an error
of roughly 10\% should be associated with his value $r_0=0.5$ fm.

Of course, in the quenched approximation, the scale
set by different quantities will differ.  But the experimental uncertainty
in $r_0$ introduces an unnecessary additional error.  
Indeed, while scales set by various directly 
measurable quantities, \ie $f_\pi$ \cite{MAYNARD}, $m_\rho$ 
\cite{UKQCD_fB00},  $f_K$ and $m_K$ \cite{LELLOUCH00}, or $m_{K^*}$
and $m_K$ \cite{APE_fB_BB00}, differ by at most 7\% at $\beta=6.2$,
the $r_0$ scale differs from these 
by as much as 15\%.
The UKQCD result $\efb=218(5)^{+\phantom{4}5}_{-41}\;\MeV$
\cite{UKQCD_fB00} becomes $195(8)^{+\phantom{4}0}_{-13}\;\MeV$ \cite{MAYNARD}
when $f_\pi$ is used to set the scale.  Other systematic errors
(some of which are discussed below) then easily account for any
remaining difference between this result and 
those of other groups.

Turning now to the CP-PACS results \cite{CPPACS_fB00,CPPACS_NEW}, we see
from Fig.~\ref{fig:fb_quenched_allgroups} that these are from
rather coarse lattices ($a\! > \! 0.57(\GeV)^{-1}$) and that the lattice
spacing dependence is quite large.  Although the Iwasaki gauge
action \cite{IWASAKI} used here 
generally gives quite small scaling violations,
decay constants are an exception \cite{BURKHALTER}.  In particular,
$f_\pi$ has an $a$-dependence that is qualitatively very similar
to that of $f_B$.  Indeed, in the heavy-clover case,
$f_B/f_\pi$ scales considerably
better than $f_B/m_\rho$ and 
gives $f_B=174(4)(3)\;\MeV$  instead of the quoted 
$188(3)(9)\;\MeV$.
Similarly, with NRQCD, $f_B/f_\pi$ gives $f_B=169(7)\;\MeV$ instead
of the quoted $191(5)(11)\;\MeV$.  In the absence of data at
smaller $a$, I believe the $f_B/f_\pi$ results are more reliable.
Furthermore,
the discrepancy suggests that the systematic error estimates
should be increased.  The CP-PACS results are then completely 
compatible with the average from other groups and do not imply
a significant increase in central value or decrease in error for quenched
$f_B$.

Before determining a world average for $f_B$, I now examine
several other sources of systematic error that affect calculations
by various groups.

\subsection{Renormalon Shadows}
\label{sec:renormalon}

In lattice HQET \cite{EICHTEN87}, 
NRQCD \cite{LEPAGE87}, and the Fermilab \cite{EKM} formalisms, 
corrections to the static limit require the addition of
higher dimensional operators, with coefficients that depend on the
heavy quark mass. Such power corrections are most easily discussed
in HQET. The framework I will use is that of
Martinelli and Sa\-chraj\-da \cite{MARTINELLIandSACHRAJDA96}; for additional
relevant discussion see Ref.~\cite{OTHERRENORMALON}.

In lattice HQET a physical quantity $\cP$, which depends on the heavy quark 
mass $m_Q$, can be expressed through order $1/m_Q$ as
\begin{equation}
\cP(m_Q)= C_1\langle|O_1|\rangle_a + \frac{C_2}{2m_Q}
\langle|O_2|\rangle_a  \ \,
\label{eq:renormalon}
\end{equation}
where the short distance coefficients $C_1$ and $C_2$ are
perturbatively calculable
functions of $am_Q$ and the coupling $g$.
The higher dimension operator $O_2$ mixes with $O_1$ with a power divergent
(like $1/a$) coefficient. 
In particular, we have
\begin{eqnarray}
C_1(am_Q,g) &=& c_1(am_Q,g) + \frac{\tilde c_1(am_Q,g)}{2am_Q}\cr
c_1(am_Q,g) &=& 1 + \# g^2 + \# g^4 + \dots\cr
\tilde c_1(am_Q,g) &=& \# g^2 + \# g^4 + \dots\cr
C_2(am_Q,g) &=& 1 + \# g^2 + \# g^4 + \dots\ ,
\end{eqnarray}
where ``$\#$'' stands for logs of $am_Q$ or constants. 
For future reference, I denote  $C_2O_2/(2m_Q)$ as
the ``bare'' $1/m_Q$ operator. Terms in $C_1O_1$ 
generated by loop corrections to the bare $1/m_Q$ operator I call
the ``subtraction.'' (The subtraction is just $[\tilde c_1/(2am_Q)]O_1$ in the 
HQET case.)  Finally, the sum of the bare $1/m_Q$ operator
and the subtraction gives
the ``renormalized'' $1/m_Q$ operator.

It is widely believed, but 
not rigorously proven, that the series $c_1$ and $\tilde c_1$
have renormalon ambiguities at $\cO(\Lambda_{\rm QCD}/m_Q)$. In the
sum $C_1$, these ambiguities should cancel. However, since the cancellation
only occurs at high order, one might anticipate \cite{MARTINELLIandSACHRAJDA96}
that the low order series for $C_1$ ``converges'' rather slowly.
I call this effect the ``renormalon shadow:'' although the renormalons
are formally gone, their influence lingers.  We expect renormalon shadows
in any lattice calculation where power law divergences are
subtracted perturbatively.

As an example, consider the lattice HQET computation
of the $b$ quark mass.
In the static limit, $m_b$ is just the meson
mass $M_B$. A non-trivial calculation must therefore include the
$1/m_Q$ correction, \ie the term of $\cO(\Lambda_{\rm QCD})$.
The relevant coefficients were calculated to two loops and combined
with lattice data in
\cite{MARTINELLIandSACHRAJDA98}. The ``bad news''
from this calculation is that the perturbative error on $m_b$
that would result from stopping at one loop is $\sim\!300\;\MeV$,
namely the size of the effect one is trying to compute. The rather
large error can be taken as evidence that the renormalon shadow
is real in this case. However,
I also find ``good news'' in 
Ref.~\cite{MARTINELLIandSACHRAJDA98}:  The $\sim\!300\;\MeV$
one-loop error could have been correctly estimated by a standard one-loop 
analysis. Such analysis involves computing the scale $q^*$ \cite{LEPMAC}
(here $q^*=1.446/a$), and seeing how the one-loop result changes under
reasonable variations in $q^*$ (say $1\le aq^* \le \pi$) and in
coupling schemes (say $\alpha_{\overline{MS}}(q^*)$ or
$\alpha_V(q^*)$).

Now let us apply what we have learned to the calculation of \fb\ 
in NRQCD.  Paralleling Eq.~(\ref{eq:renormalon}),
the NRQCD expression for \fb\  takes roughly the form
\begin{equation}
\efb \propto C_1\langle0|O_1|B\rangle_{a,m_Q} + \frac{C_2}{2m_Q}
\langle0|O_2|B\rangle_{a,m_Q}\ ,
\label{eq:fbNRQCD}
\end{equation}
where
\begin{equation}
O_1=\bar q\gamma_0\gamma_5 Q\ ;\qquad O_2=
\bar q\gamma_0\gamma_5 \Dslash_\perp Q
\ ,
\label{eq:O1-2def}
\end{equation}
with $Q$ the heavy quark field, $q$ the light quark field and
$\Dslash_\perp \equiv \sum_{i=1}^3 \gamma_i D_i$ the spatial
Dirac operator. Note that, unlike in
Eq.~(\ref{eq:renormalon}), the matrix elements in Eq.~(\ref{eq:fbNRQCD}) 
depend explicitly on $m_Q$.  This is because $m_Q$ already appears in the
lowest order NRQCD Lagrangian. 

As in the HQET case,
$O_2$ has a $1/a$ divergence proportional
to $O_1$.\footnote{ 
Note, however, that
the separation of
$C_1$ into $c_1$ and $\tilde c_1$ is now no longer  meaningful
since the since the loop diagrams can
generate arbitrary functions of $aM$.}
Therefore one expects that a renormalon shadow will appear, producing large
errors in the
renormalized $1/m_Q$ matrix element unless the perturbative calculation
is taken to high order. However,
because the matrix element of $O_1$
also depends on $m_Q$,  
the renormalized $O_2$ matrix element in NRQCD
is not the entire $\cO(1/m_Q)$ effect.
This is a crucial difference from the HQET case.
The importance of the renormalon shadow here
is thus a numerical question.

The discussion of \fb\  in the Fermilab formalism is similar.
In Eq.~(\ref{eq:fbNRQCD}), one just replaces
$1/(2m_Q)$ by $ad_1$, where $d_1$ is the coefficient of the ``rotation''
defined in \cite{EKM} and is a function of
$am_Q$.  As $am_Q\to\infty$, $d_1\to 1/(2am_Q)$, thereby
reproducing NRQCD.
Other the other hand, $d_1\to0$ as $a\to0$.  This means
that the subtraction here does not diverge
(and in fact vanishes) as $a\to0$.
Note however that the relative perturbative uncertainty in the
renormalized $1/m_Q$ operator is unaffected by  the presence of
$ad_1$, which is just an overall factor.
$O_2$ itself is still power divergent
and a renormalon shadow
should still appear.  Still, the factor $ad_1$ may reduce the
numerical
importance of the shadow.

To discuss the renormalon shadow in \fb\  quantitatively, I start with
the Fermilab formalism, using
MILC data \cite{MILC_LAT00} at $\beta=6.15$ with the 
non-perturbative \cite{ALPHA_CSW-CA-ZA} value of the 
clover coefficient $C_{SW}$. The perturbative corrections  
have been calculated \cite{IOY} at one loop.  I define the subtraction
here as $O_1$ times the difference between its complete
perturbative coefficient and the coefficient when $O_2$ is omitted
from the calculation \cite{ISHIKAWA}.  Since $q^*$ has not been
computed for this quantity, I choose the value that
comes from the static-light $Z_A$ \cite{CB-TD}
($q^*=2.85/a$)  and use $\alpha_V(q^*)$ 
\cite{LEPMAC}.
To get the uncertainty
in the subtraction, I then replace $q^*\to 1/a$. (The change is
smaller when $q^*$ is  increased, even to $10/a$, or when the scheme
is changed to $\overline{\rm MS}$ or ``boosted perturbation theory''
\cite{BOOSTEDPT}.)
The ratio
of the uncertainty to the renormalized $1/m_Q$ matrix element is
then quite large: $60\%$.  I take this as evidence that a renormalon
shadow is indeed present.

\begin{figure}[tb]
\vspace{-0.4truein}
\includegraphics[bb = 100  200 4096 4196,
width=2.7truein,height=2.3truein]{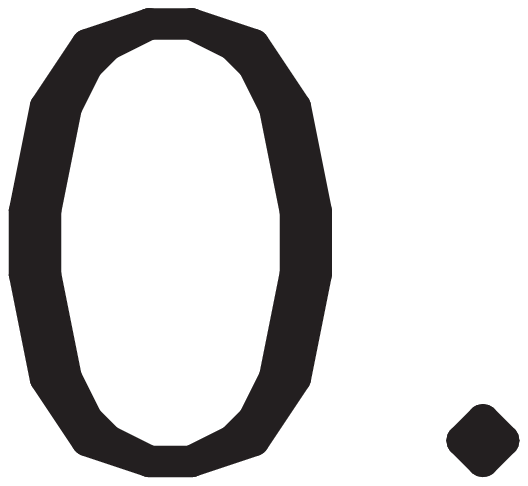}
\vskip -.15truein
\caption{}{Effect of bare and renormalized $1/m_Q$ operator on heavy-light
decay constants in the Fermilab formalism. $M_{HL}$ is the mass
of a heavy-light pseudoscalar meson, and $f_{HL}$ is its
decay constant.  MILC data ($\beta=6.15$, $C_{SW}=
1.644$) \cite{MILC_LAT00} is used.}
\label{fig:Fermilab_shadow}
\vskip -.1truein
\end{figure}

However, since the renormalized $1/m_Q$ operator contributes only a small
amount to \fb, the ultimate effect of the the renormalon shadow
is small.  Figure \ref{fig:Fermilab_shadow} shows the separate
effects of the bare $1/m_Q$ operator and the subtraction, as well
their combined effects (the renormalized $1/m_Q$ operator).  Clearly,
most of $f_B$, as well as its $m_Q$ dependence, comes from the
matrix element of $O_1$. The matrix element
of the bare $1/m_Q$ operator adds $4.6\%$ to \fb; while the subtraction
is $-2.6\%$.  
The renormalized $1/m_Q$ operator thus
contributes only $(2.0\pm 1.2)\%$, where I've included
the uncertainty from the renormalon shadow.  The uncertainty
($\sim\! 2\;\MeV$) is smaller than many other systematic effects
in current computations. Note that the error
which would be made from including the bare $1/m_Q$ operator without
the one-loop subtraction is probably larger than leaving out the
operator completely.

In the NRQCD case, most of the numerics (for \fbs\  at $\beta=6.0$) can be found
in Ref.~\cite{COLLINS00}. Here the bare $1/m_Q$ contributes $\sim\! 11\%$;
the one-loop subtraction, $\sim\!-7\%$.  The inclusion of the
renormalized $1/m_Q$ operator is thus a $(4\pm 4)\%$ effect, where
the error again comes from my varying $q^*$ in the subtraction.  
It is no surprise that
the effect and error here are larger than in the Fermilab case: 
$1/(2am_Q)$ 
is considerably larger than 
$d_1$
for relevant values of $am_Q$.
However, the $4\%$ uncertainty ($\sim 7\;\MeV$) is still smaller
than several other systematic effects.

My conclusion is that the calculation of leptonic decay constants
of B mesons is under control for both the Fermilab formalism
and NRQCD, despite the presence of renormalon shadows.  The issue however
needs to be considered on a case-by-case basis.  There is no
guarantee that renormalon shadows are negligible for other physical quantities, 
systems, or orders in perturbation theory.  Indeed, in many
computations the perturbation theory is only put in at (tadpole-improved)
tree level.  The subtractions 
are therefore left out entirely.  From the above discussion it seems likely
that the inclusion of higher dimension operators in such a situation
is worse them omitting them entirely, and may lead to large
renormalon shadow uncertainties (see Sec.~\ref{sec:spectra}).

Two further comments: (1) My numerical estimates of
renormalon shadow uncertainties 
are completely standard and do not involve the details
of the renormalons at all.  Therefore
it is possible 
to take the term ``renormalon shadow'' as merely a convenient
shorthand for the phrase ``large perturbative uncertainties 
in power divergent subtractions.''  (2) The error I estimate
for renormalon shadows is of course
not the only perturbative uncertainty: many of the terms in $C_1$
would be present even if $O_2$ were omitted. Once it is
ascertained that renormalon shadows are under control, it is probably 
preferable to quote simply an overall perturbative uncertainty,
obtained by varying $q^*$ everywhere. At the current state of the art
(one loop) the overall perturbative uncertainty is generally 
larger than what I estimated above for the renormalon shadow.  
At high enough order, my method would probably overestimate
the perturbative uncertainty 
because I have not taken into account the expected
cancellation of renormalons between $c_1$- and $\tilde c_1$-type terms.

\subsection{Discretization Errors in MILC Data}

In MILC calculations of quenched $f_B$ \cite{MILC_PRL} using Wilson
and static heavy quarks, the most important source of systematic error
was the continuum extrapolation.  This error was manifest
in the difference between an linear extrapolation of
\fb\  from all data sets ($0.2\; \GeV^{-1}\ltwid a \ltwid 0.8\; \GeV^{-1}$) and 
a constant extrapolation from only the finer lattices 
($a\ltwid0.5\;\GeV^{-1}$; $\beta\ge6.0$).  As of June, 1999 (with additional
running from what appeared in \cite{MILC_PRL}),
the former extrapolation gave $\efb=154(11)\;\MeV$; while that latter,
$180(10)\;\MeV$.  On the basis of the behavior of
\fbsofb, we chose the linear extrapolation for the
central value and  took the large $26\;\MeV$ difference as the
discretization error.

Since that time,  two developments have significantly reduced the
discretization error we quote.  First of all, Tom DeGrand and I
have recalculated \cite{CB-TD} the one-loop scale $q^*$ appropriate
to the static-light axial current 
renormalization constant $Z^{SL}_A$.  Instead of Hernandez and Hill's result
$q^*=2.18/a$ \cite{HandH} for tadpole-improved light Wilson fermions,  
we find a $q^*$ that is mildly dependent on $am_Q$
(as $Z^{SL}_A$ is) and is
$\approx\! 1.4/a$ for typical values of $am_Q$.  We differ from Ref.~\cite{HandH}
because: (a) we define $q^*$ using the complete integrand
for $Z^{SL}_A$ (including the
continuum part, which gives the $m_Q$ dependence),  
and (b) we do not discard pieces of the $Z^{SL}_A$ integrand which vanish
by contour integration --- such pieces do not vanish for $q^*$ because
of the additional $\ln(q^2)$ factor.  Difference (b) is responsible for
most of the discrepancy.

In the MILC calculation, the static-light $q^*$ described above is employed as
the central value of $q^*$ not only in the static-light renormalization
but also in the (propagating) heavy-light renormalization, for which
$Z_A$, but not $q^*$, has been computed \cite{KURAMASHI}.  The new
value $q^*\approx 1.4/a$ considerably reduces the $a$-dependence of \fb\  
with Wilson quarks \cite{MILC_LAT00}.  
The difference between
the two extrapolations of this data is now  $15\;\MeV$ instead of 
$26\;\MeV$ (see Fig.~\ref{fig:fb_quenched_MILC}).

\begin{figure}[tb]
\null
\vspace{-.1truein}
\includegraphics[bb = 100  200 4096 4196,
width=2.7truein,height=2.6truein]{fb_quenched_MILC.ps}
\vskip -.1truein
\caption{}{Updated MILC data \cite{MILC_LAT00} for quenched $f_B$ {\it vs.}\/
lattice spacing. Wilson and nonperturbative
clover (``NP'') fermions  with the Fermilab formalism are used.
 ``NP-tad'' and ``NP-IOY'' represent two different
ways of renormalizing the heavy-light axial current --- see text.  }
\label{fig:fb_quenched_MILC}
\end{figure}

In addition to reanalyzing the Wilson data,
we recently completed 
running at $\beta=6.0$ and $6.15$ with nonperturbative 
\cite{ALPHA_CSW-CA-ZA} 
clover fermions and the full
Fermilab formalism through order $1/m_Q$.
The $1/m_Q$ operator is always renormalized at one loop \cite{IOY,ISHIKAWA}, as in 
Sec.\ \ref{sec:renormalon}.  For the rest of the renormalization
of the heavy-light axial current, we use either the one-loop calculation
\cite{IOY}, or a ``tadpole renormalization'' Ansatz designed
to reproduce the nonperturbative renormalization of Ref.~\cite{ALPHA_CSW-CA-ZA}
at small mass and have a sensible limit as $m_Q\to\infty$.  These two
approaches are shown as ``NP-IOY'' and ``NP-tad,'' respectively,
in Fig.~\ref{fig:fb_quenched_MILC}.  NP-tad is described in more
detail in Sec.\ \ref{sec:MILCvsUKQCDnorm}.  Both
NP-IOY and NP-tad are correct to $\cO(a)$.  Higher order effects
(\eg $\cO(a^2)$, $\cO(g^4)$) are complicated but, hopefully, rather
small.  Each approach is fitted to a constant; the difference in extrapolated
value from the two NP approaches gives an estimate of these higher order
effects.

From Fig.~\ref{fig:fb_quenched_MILC} it now seems clear that it was a mistake
to choose only the
linear extrapolation to find the central value of the Wilson
data in the continuum:
the linear fit gives the 
lowest extrapolated value of all four approaches. (In the case of \fbsofb,
the linear extrapolation differs even more from the other three fits.)
MILC currently averages all four approaches for the central value
and defines the discretization error as the standard deviation of the four.
The result 
appears as the quenched ``MILC00'' 
point in Fig.~\ref{fig:compare_fb} and Table~\ref{tab:fb}.

\subsection{``Nonperturbative'' Heavy-Lights }
\label{sec:MILCvsUKQCDnorm}

APE \cite{APE_fB_BB00,BECIREVIC00}, UKQCD \cite{UKQCD_fB00,MAYNARD}, and 
MILC \cite{MILC_LAT00}
have simulated heavy-light physics using nonperturbative clover
fermions.
The starting point 
 is the expression for the improved, renormalized
axial current $A_0^R$:
\begin{eqnarray}
A_0^R\!\!\! &=&\!\!\!\!\! Z_A\sqrt{4\kappa_q\kappa_Q}
(1\! + \!\frac{b_Aam_{Q,0}}{2}) 
[A_0\!\! +\!\! c_Aa\partial_0P_5
]\ ,\cr
A_0\!\!\! &=& \!\!\!\bar q\gamma_0 \gamma_5 Q\ ; \qquad 
P_5 = \bar q \gamma_5 Q \ ,
\label{eq:A_R}
\end{eqnarray}
where $\kappa_q,\ \kappa_Q$ are the light  and heavy quark hopping
parameters, respectively, and $m_{Q,0}$ is the 
bare heavy quark mass ($am_{Q,0}\equiv 1/(2\kappa)-1/(2\kappa_c)$).
For simplicity, the light quark mass has been set to zero.

All three groups use the nonperturbative values of
$Z_A$ and $c_A$ in Eq.~(\ref{eq:A_R}) 
(as well as  the  clover coefficient $C_{SW}$)
computed by the 
Alpha collaboration \cite{ALPHA_CSW-CA-ZA}. 
The groups differ, however, on the choice of the
coefficient $b_A$, which was not computed nonperturbatively
in Ref.~\cite{ALPHA_CSW-CA-ZA}. At $\beta=6.2$, for example, 
APE uses $b_A=1.24$ 
from the one-loop calculation \cite{SINTandWEISZ} and
boosted perturbation theory \cite{BOOSTEDPT}; 
while UKQCD uses $b_A=1.47$ from a preliminary
nonperturbative calculation of Bhattacharya \et\ \cite{BHATTACHARYA99}.
This difference accounts for any disagreement between APE
and UKQCD results that remains once the scales are set in the same way.
MILC's $b_A$ is taken from perturbation theory \cite{SINTandWEISZ}, but
with coupling $\alpha_V(q^*)$, with $q^*$ chosen as the value ($\approx\!1/a$)
which produces the nonperturbative result \cite{ALPHA_CSW-CA-ZA} for the
similar quantity $b_V$.  This gives values of $b_A$ ($1.47$ at $\beta=6.15$
and $1.42$ at $\beta=6.0$) that are quite close to those 
of UKQCD.\footnote{At $\beta=6.0$ and $6.2$, a new nonperturbative calculation 
of all the coefficients, including
$b_A$, is now available \cite{BHATTACHARYA00}. It will be interesting
to see how these values affect the results for $f_B$.}

APE and UKQCD  apply Eq.~(\ref{eq:A_R}) directly
for moderate values of $am_{Q,0}$ 
(up to $\sim\!0.5$ at $\beta=6.2$ and $\sim\!0.75$ 
at $\beta=6.0$). This allows them to reach a maximum meson
(pole) mass of $\approx\!2\;\GeV$ at $\beta=6.2$, where I've set
the scale by $m_\rho$.
Using HQET, which implies that $f_{HL}\sqrt{M_{HL}}$ should be
a polynomial in $1/M_{HL}$ (up to logs), they then attempt 
to extrapolate the results up to the $B$ mass. ($M_{HL}$ and $f_{HL}$ 
 are the mass and decay constant of a generic heavy-light
pseudoscalar meson.)

There are of course systematic errors associated with this
approach. First of all, one does not know {\it a priori} what
order polynomial to use; there is a large difference ($\sim\!\!10\%$)
between the results of linear and quadratic extrapolations.
A second systematic effect is more subtle.  
Although discretization errors in Eq.~(\ref{eq:A_R}) 
are very small for $am_{Q,0}\ll 1$ and appear to remain quite small
even for the maximum $am_{Q,0}$ used by ULQCD and APE, those errors
grow rapidly with $am_{Q,0}$.  Indeed, for $am_{Q,0}\to\infty$, 
$A_R$ in Eq.~(\ref{eq:A_R}) does not approach a static limit
as it should, but goes to $-\infty$ because $c_A<0$ and $\partial_0P_5
\sim \sinh (M_{HL,1}) \sim am_{Q,0}$. (Here
$M_{HL,1}$ is the meson pole mass.) Even if $c_A$ were zero, $A_R$
would still blow up because of the term $b_Aam_{Q,0}$.
Thus, small discretization errors may be magnified significantly
by the extrapolation to the $B$.

To estimate the latter error, one may compare to the MILC ``NP-tad''
approach
\cite{THANKSINT}.
The MILC goal is to
replace Eq.~(\ref{eq:A_R}) by an expression that is equivalent through
$\cO(a)$ but which gives a sensible limit for $f_{HL}\sqrt{M_{HL}}$ 
as $am_{Q,0}\to\infty$. The result is then used at arbitrary
$am_{Q,0}$, {\it \`a la} Fermilab.   To do so, we first define
\begin{equation}
\label{eq:Rdef}
R(M_{HL})\equiv \frac {\langle 0|\partial_0P_5|HL \rangle}
{m_{Q,0}\langle 0|A_0|HL \rangle}\ ,
\end{equation}
where $HL$ is a generic heavy-light pseudoscalar meson.
Due to a cancellation of $\sinh (M_{HL,1})$ (from $\partial_0$)
and the explicit $m_{Q,0}$ in the denominator, one expects $R$ has
a finite limit as $am_{Q,0}\to\infty$.  This is confirmed by simulations.
Then
\begin{equation}
\label{eq:MILC_A_R}
A_0^R=Z_A\sqrt{4\kappa_q\kappa_Q}
\sqrt{1 + (b_A+2c_AR)am_{Q,0}} 
A_0
\end{equation}
gives results 
for $\langle 0|A_0^R|HL \rangle$ that are
identical to Eq.~(\ref{eq:A_R}) through $\cO(a)$.
However,
because $\kappa_Qam_{Q,0}\to 1/2$ as $\kappa_Q\to0$,
Eq.~(\ref{eq:MILC_A_R}) has a static limit, unlike (\ref{eq:A_R}).
Indeed, (\ref{eq:MILC_A_R}) is just a version of the
Fermilab formalism at tadpole-improved tree level (the $1/m_Q$
operator is omitted for simplicity), but with a special
(mass-dependent) value for the tadpole factor: $u_0 = (b_A+2c_AR)^{-1}$.
The similarity to tadpole improvement (within the context
of nonperturbative renormalization) is the reason for the
name ``NP-tad.''

\begin{figure}[tb]
\null
\vspace{-.5truein}
\includegraphics[bb = 100  200 4096 4196,
width=2.7truein,height=2.7truein]{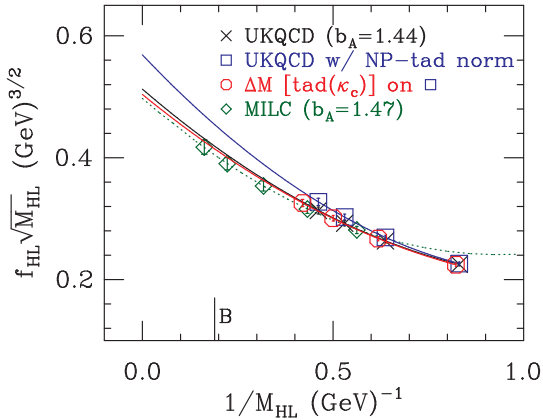}
\vskip -.1truein
\caption{Effects, at $\beta=6.0$, of various treatments of heavy-lights using
nonperturbative clover fermions. UKQCD results (crosses) have
been
adjusted to same scale (from $f_\pi$) as MILC results (diamonds).  
Squares show the effect of adjusting the crosses using the NP-tad 
normalization.  Octagons include the tadpole-improved
mass shift (described in text) 
on the squares.
}
\label{fig:NPHL}
\vskip -.05truein
\end{figure}

Figure \ref{fig:NPHL} shows the effect of reanalyzing the UKQCD results
with the NP-tad normalization.  This
$\cO(a^2)$ effect is $2\%$ to $5.5\%$ on their data points, but
$8.5\%$ after extrapolation to the $B$.

An additional difference between MILC and UKQCD or APE is that MILC,
following the Fermilab formalism, defines the meson mass as the
kinetic mass $M_2$, rather than the pole mass $M_1$.  Since the difference
in masses is $\cO(a^2)$ this does not affect the $\cO(a)$ improvement.
$M_2$ is determined
by adjusting the measured meson
pole mass upward by $\Delta M \equiv m_2-m_1$,
where $m_2$ and $m_1$ are the heavy quark
kinetic and pole masses, respectively, 
as calculated  in tadpole-improved tree approximation \cite{EKM},
with the mean link fixed by $\kappa_c$.  Figure~\ref{fig:NPHL} 
shows the effect of this shift on the UKQCD data (after first changing
to NP-tad normalization).  
One can check the MILC determination of $M_2$ by using instead
the $M_2$ values directly computed by
UKQCD from the meson dispersion relation.  The resulting curve is
not shown in Fig.~\ref{fig:NPHL} because it is indistinguishable
from the fit to the diamonds.

Note that the effects of the NP-tad norm and the shift by $\Delta M$
almost cancel, so that the final results of UKQCD and MILC
are actually quite consistent as long as the scale is set in the
same way.  Since the two effects are logically  independent, however,
the $17\;\MeV$ difference for $f_B$ among the  curves
in Fig.~\ref{fig:NPHL} is a measure of the discretization error.  
This error is still $15\;\MeV$ at $\beta=6.2$. 
I would not, however, argue that this error should be added
to those quoted by UKQCD.  They already include a discretization error
error obtained by comparing the results at $\beta=6.2$ and $6.0$.
Indeed, the discretization error computed that way is $16\;\MeV$
(using an $m_\rho$ scale), identical to the estimate above.

The goal of the approaches discussed in this subsection is
to treat the heavy-lights in a way that
takes advantage of the known nonperturbative renormalizations for
light-lights.  But since one either works in or extrapolates to a 
region where $am_Q$ is not small, the systematic errors
are not necessarily small.  Furthermore, because the methods agree
at $\cO(a)$, there is
no way {\it a priori} to distiguish among them at the level of
Eq.~(\ref{eq:A_R}). If one insists on using the nonperturbative
information for the heavy-lights, then I prefer the NP-tad approach 
(and equating the physical mass with $M_2$) because (a) having
a static limit seems desirable if we want to use (or extrapolate
to) large $am_Q$, and (b) it appears to scale somewhat better:
the difference between \fb\ at $\beta=6.0$ and $6.15$ is only
$2\;\MeV$ (see Fig.~\ref{fig:fb_quenched_MILC}).  However, this
is far from definitive: point (a) is subjective,  and point (b) is
not very convincing with only two lattice spacings.

The situation will improve soon.  Although a true nonperturbative
treatment of heavy-lights seems difficult, a nonperturbative
computation of the $\cO(a)$ renormalization/improvement coefficients
for the {\it static}-light axial current is close to completion \cite{SOMMER00}.
This will be an important advance, since it will allow one to compute
$f_B$ with an interpolation between two nonperturbative calculations.
The extrapolation error of the UKQCD or APE approach will therefore
be much reduced.  One will also be able to see how well NP-tad
is really doing in the intermediate region.

A alternative computation with nonperturbative clover
fermions that can be done today is the
``NP-IOY'' approach mentioned
in the previous section. However, since it is just a
marriage of standard techniques, it is perhaps misleading
to give it a special name: One simply
confines the use of
the nonperturbative information to the light-light sector,
where it is completely justified.  In this case that means setting
the scale with, for example, $f_\pi$ or $m_\rho$. 
The straightforward Fermilab approach, with one loop perturbation
theory \cite{IOY}, is then used for the heavy-lights.

It is also worth mentioning here the calculation of Ref.~\cite{LELLOUCH00}.
This is similar to Refs.~\cite{APE_fB_BB00,UKQCD_fB00} in that the B meson
is reached by extrapolation from relatively small values of $am_Q$.
However the improvement is done perturbatively, and the EKM normalization
\cite{EKM}
(but not the shift $M_1\to M_2$) is used for the central values.
I believe that the errors quoted in Ref.~\cite{LELLOUCH00} adequately
account for all systematic effects (within the quenched approximation) 
of that calculation.

\subsection{Unquenching \fb}
Figure~\ref{fig:fb_dynamical_allgroups} shows the current status
of $f_B$ calculations with two flavors ($N_f=2$) of dynamical (sea) fermions.
One improvement over last year is that the very low value from 
the MILC ``fat-link'' clover computation \cite{MILC_LAT99} 
has now been understood.
From measurements of the quenched static quark potential,
we have shown that fattening
smooths out the potential well at short distances
relative to the corresponding
``thin-link'' (standard) case.  Indeed, for the 
fattening MILC used (10 iterations of APE smearing \cite{APESMEAR}
with staple coefficient $\alpha=0.45$), the potential at distance 1
(for $\beta=5.85$)
is increased by $45\%$.  At distance $\sqrt{3}$, the increase is only
to $8\%$.  And by distance 3, it is under $1\%$.  Such an effect
is not surprising; the expected range for this amount of fattening is
$\sim\!\sqrt{10\times4.5}\approx2.1$ lattice spacings \cite{CB-TD_LAT99}.
One expects that when the potential well at short distance becomes
shallower, \fb\ decreases, since it is basically the wave function at
the origin. 

\begin{figure}[tb]
\null
\vskip -.15truein
\includegraphics[bb = 100  600 4096 4196,
width=2.5truein,height=2.3truein]{fb_dynamical_allgroups.ps}
\caption{}{World data for $f_B$ with $N_f=2$ {\it vs.}\/ 
lattice spacing. Only statistical errors are shown.
``MILC99'' is Ref.~\cite{MILC_LAT99};
see Table~\ref{tab:fb} for other references.}
\label{fig:fb_dynamical_allgroups}
\end{figure}

On quenched configurations
at $\beta=6.0$, $f_B^{\rm fat}/f_B^{\rm thin} = 0.76$ when ``thin'' is
the NP-tad approach, and $0.73$ when ``thin'' is NP-IOY.
The extent of the effect was somewhat surprising to us, 
since the light hadron spectrum behaves quite well with fat-link
clover fermions \cite{STEPHENSON} and since the potential is only changed
by an amount $>\! 10\%$  for distances $<\! 0.16$ fm.  The difference
between fat and thin may be exaggerated by the fact that in the fat case
the one-loop renormalization of heavy-light $Z_A$ has not been
computed and the static-light
$Z_A$ \cite{CB-TD_LAT99,CB-TD} was used instead.  Despite this,
the ratio $f_B^{\rm fat}/f_B^{\rm thin}$ should be roughly the same
at $N_f=2,am=0.01,\beta=5.6$ (the parameters MILC used in the unquenched
case) as at quenched $\beta=6.0$ because the lattice spacings
and couplings are quite close.  Under this assumption, I plot
$f_{B,N_f=2}^{\rm fat,corrected} = f_{B,N_f=2}^{\rm fat} \times
f_{B,quench}^{\rm thin}/f_{B,quench}^{\rm fat}$ in 
Fig.~\ref{fig:fb_dynamical_allgroups}, with thin=NP-tad. 

The MILC result for $f_{B,N_f=2}$ in the continuum is 
obtained by extrapolating
the Wilson data under two different assumptions that seem likely
to bracket the real behavior:
(1) constant behavior as $a\to0$,  and (2) a linear decrease from smallest
current $a$ values ($\approx \! 0.45\;\GeV^{-1}$) with a slope equal
to that of the linear fit to the quenched data 
(see Fig.~\ref{fig:fb_quenched_MILC}).   The average of the two
procedures is taken as the central value; the standard deviation, as the
extrapolation error.  The (preliminary) result shown in 
Fig.~\ref{fig:compare_fb} and Table~\ref{tab:fb} then is:
$f_B =  191(6)({}^{+24}_{-18})({}^{+11}_{-0})\;\MeV$, where the
errors are respectively statistical, systematic 
within the $N_f=2$ approximation,
and an estimate of the error of neglecting the strange sea quark.
Although the consistency of the corrected fat-link results 
with the Wilson data is comforting, the former
is not included in MILC final results
because there is no guarantee that the correction factor is
the same in the quenched and full cases.

As in the quenched case,
the preliminary CP-PACS results for \fb\ with $N_f=2$ use the
Iwasaki gauge action \cite{IWASAKI}. They find \cite{CPPACS_fB00,CPPACS_NEW}:
$f_{B,N_f=2}= 208(10)(11)\;\MeV$ with the Fermilab
formalism (heavy clover) and $204(8)(15)(+44)\;\MeV$ with NRQCD; these
results are shown in Fig.~\ref{fig:compare_fb}.
The smallest-$a$
points in Fig.~\ref{fig:fb_dynamical_allgroups}
are taken as the ``continuum'' values, and the discretization
error comes theoretical estimates: $\cO(a^2\Lambda^2)$ and
$\cO(\alpha a\Lambda)$ in the heavy-clover case.
However, my comment about discretization errors in the quenched
CP-PACS data applies here too.  Indeed, comparing 
Fig.~\ref{fig:fb_dynamical_allgroups} with Fig.~\ref{fig:fb_quenched_allgroups},
one sees that the lattice spacing dependence of the $N_f=2$  data
is if anything greater than in the quenched case.  The theoretical
error estimates do not account for the full amount of
the variation with $a$. I therefore
think the procedure for finding central values and errors is overly
optimistic.  

My own rough estimate from a $const\!+\!a^2$ fit
to the CP-PACS heavy-clover data
would be $f_{B,N_f=2}= 190(12)(26)\;\MeV$. The systematic error in
my estimate is dominated by the discretization error, taken from the
difference of the extrapolation and the smallest-$a$ point.  Although
one can certainly argue with my extrapolation (there are 
for example $\alpha a \Lambda$ errors in addition to $a^2$ errors)
I believe the systematic error of $\sim\!26\;\MeV$ is reasonable.

One difference between CP-PACS and MILC data shown 
in Fig.~\ref{fig:fb_dynamical_allgroups} is that the MILC data
is ``partially quenched'' --- the sea quark mass is held
fixed while the valence mass is extrapolated to the
physical $u$,$d$ mass.  The CP-PACS points are fully unquenched,
with the valence and sea masses extrapolated together.  However,
MILC has repeated the analysis in a fully unquenched manner;
the points are then still roughly constant in $a$, and the final result
is changed only slightly.  The difference is included in the
MILC systematic error.

In the NRQCD $N_f=2$ points in Fig.~\ref{fig:compare_fb} 
\cite{COLLINS99,CPPACS_fB00}, the systematic errors are quite large.
The dominant source of these errors is the $\sim\!25\%$
difference in the lattice scale from that set by light
quark physics ($m_\rho$, say) and the 1P-1S $\Upsilon$
splitting.  In previous, quenched simulations, it was generally assumed
that the scale difference was due to quenching and would
go away with dynamical quarks.  Although the difference is
in fact somewhat smaller for $N_f=2$ than in the quenched approximation,
it is still large; there is no indication that it would vanish
in the physical case, \ie 3 light dynamical quarks.  Perhaps
the errors for heavy-heavy spectra in NRQCD
are underestimated (see Sec.\ \ref{sec:spectra}). 
I do not, however, have the data and calculations to subject this suspicion to a
quantitative test.

Looking at Fig.~\ref{fig:compare_fb} and recalling my remarks on
discretization effects, one sees that the systematic errors
on \fb\ in both the quenched and the 2-dynamical-flavor cases are still
rather large overall.  However, comparing quenched and $N_f=2$
results by the same groups, for which many of the systematic effects
cancel, one can say with some confidence that $f_{B,N_f=2}$ is
about $10$--$15\%$ larger than $f_{B,quench}$.  The effect
of unquenching is roughly the same for \fbs;
while the increase
for \fd\ and \fds\ appears somewhat smaller, about $3$--$8\%$ 
\cite{MILC_LAT99,MILC_LAT00,CPPACS_fB00,CPPACS_NEW}.

Taking into account all above remarks, I arrive at the
following world averages:
\begin{eqnarray}
f_{B,quench} =  175(20)\;\MeV,&&
f_{B}  =  200(30)\;\MeV,\cr
\left(\frac{f_{B_s}}{f_B}\right)_{\!quench} =
1.15(4),&&
\frac{f_{B_s}}{f_B}=
1.16(4),\cr
f_{D_s} =  255(30)\;\MeV,\phantom{junk}&&
\label{eq:fBresults}
\end{eqnarray}
where quantities without the ``$quench$'' subscript are supposed
to be full QCD quantities, including an estimate of the effect (and error)
that would result from including the strange sea quark.
Starting with the
quenched results  and adjusting them by the expected increase in the full theory
would result in similar unquenched values
and errors.

\section{B PARAMETERS}
\label{sec:bparam}

\begin{figure}[tb]
\null
\vspace{-.25truein}
\includegraphics[bb = 100  200 4096 4196,
width=2.8truein,height=2.7truein]{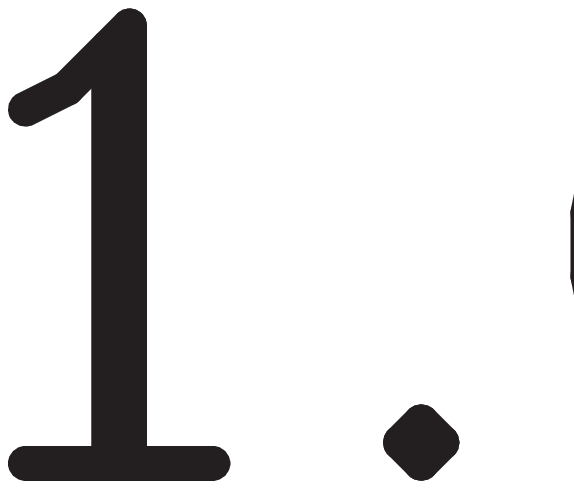}
\vskip -.15truein
\caption{}{Selection of recent data for 
quenched $\hat B_{P_d}$ {\it vs.} $1/M_{HL}$,
where $P$ is a general heavy-light pseudoscalar and $M_{HL}$ is its mass.
Only statistical errors are shown for
heavy-light data and extrapolations 
\cite{LELLOUCH00,APE_fB_BB00,BECIREVIC00}.   Static-light 
points \cite{GIMENEZandREYES,CDM} include perturbative
systematic errors. 
JLQCD \cite{JLQCD_BB00,JLQCD_BBperturb} uses NRQCD-light; 
the 4 points at each mass represent different
ways of treating the  $\cO(\alpha_s^2)$  and $\cO(\alpha_s p/m_Q)$ corrections.}
\label{fig:BB_allgroups}
\end{figure}

Figure~\ref{fig:BB_allgroups} presents most of the recent world 
data on the \bbbar\
mixing parameter \hbbd, where the ``hat'' indicates the renormalization
group invariant quantity at NLO.  Three types of calculations are shown: 
heavy-light with relativistic heavy quarks (here, clover quarks)
extrapolated to the $B$ meson mass 
\cite{APE_fB_BB00,BECIREVIC00,LELLOUCH00}, static-light 
\cite{GIMENEZandREYES,CDM},
and NRQCD-light \cite{JLQCD_BB00,JLQCD_BBperturb}.

In the relativistic heavy-light case, the extrapolation to the $B$ is motivated
by the heavy quark effective theory. HQET implies 
that $B_{P_d}$ is a power series (up to
logs) in $1/M_{HL}$, where $B_{P_d}$ is 
the (left-left) B parameter of a heavy-light
meson $P$ (mass $M_{HL}$) with a physical $d$ 
quark and an arbitrary heavy quark.
Both APE \cite{APE_fB_BB00} and Lellouch and Lin \cite{LELLOUCH00} make linear
fits in $1/M_{HL}$, resulting in the extrapolated points shown in Fig.~\ref{fig:BB_allgroups}.
Although it must be admitted that the data of Ref.~\cite{LELLOUCH00}
fit the linear form quite well, I have concerns about trusting
the HQET down to $M_{HL}\approx 1.2 GeV$ 
($1/M_{HL}\approx 0.85\;\GeV^{-1}$), which is the lightest mass
shown in Fig.~\ref{fig:BB_allgroups}.  Indeed, since the relativistic
points in Refs.~\cite{APE_fB_BB00,LELLOUCH00} suffer from some 
of the same systematic
effects discussed in Sec.\ \ref{sec:MILCvsUKQCDnorm} (\eg 
the choice of $M_1$ or $ M_2$),
the  extremely linear behavior may well be an accident.  I therefore will
omit the $0.85\;\GeV^{-1}$ point from my analysis.

References \cite{APE_fB_BB00,LELLOUCH00} do not take into account 
the static-light
results (\eg Refs.~\cite{GIMENEZandREYES,CDM}) in their extrapolations.
APE justifies this by noting that the perturbative error on the
static-light B parameter is large, while the heavy-light points in
\cite{APE_fB_BB00} use nonperturbative normalization. In my view, however,  the
$am_Q\!\sim\!1$ systematic errors in the heavy-light case 
(sec.~\ref{sec:MILCvsUKQCDnorm}) are comparable to the perturbative
errors of the static-light method. Therefore, I choose to make
a quadratic fit (in $1/M_{HL}$) to the APE points as well as the
Gimenez and Reyes/UKQCD static-light point. (The latter probably has 
the smallest systematic errors of the available static-light data,
since it uses improved light quarks at the weakest coupling, $\beta=6.2$.)
My fit is shown in Fig.~\ref{fig:BB_allgroups}. At the B mass, it gives
$\ehbbd=1.30(6)$ (statistical error only).  
Note that the extension of the fit comes reasonably
close to the $0.85\;\GeV^{-1}$ point; alternatively, including
this point in the fit would produce only a $1\%$ change in \hbbd.

An important advance in the past year is the 
calculation \cite{JLQCD_BBperturb}
of the one-loop renormalization of the $\Delta B\!=\!2$ 4-quark
operators in the NRQCD-light case.  Previously \cite{JLQCD_BB99},
the static-light renormalization constants had to be used
instead, and the result for \hbbd\ was $\sim\!\!20\%$
low compared to the extrapolated values of Refs.~\cite{APE_fB_BB00,LELLOUCH00}
and still  $\sim\!15\%$ below my interpolated value.
(Note, however, that the perturbative error in  Ref.~\cite{JLQCD_BB99}
was estimated rather realistically as
$\sim\!10\%$.)
The current perturbative calculation includes
terms of $\cO(\alpha_s/(am_Q))$, but not those of
$\cO(\alpha_s p/m_Q)$, where $p$ is the typical quark 3-momentum
in the B meson. The various squares at the same values of
$1/M_{HL}$ in Fig.~\ref{fig:BB_allgroups} represent an attempt
to estimate the systematic errors of $\cO(\alpha_s p/m_Q)$
as well as $\cO(\alpha_s^2)$.  The results are 
consistent with those of my interpolation. Unfortunately, I do not 
have the data to 
perform a renormalon shadow analysis on the NRQCD-light results,
but my guess is that the effect is no larger than in the $f_B$ case.

Taking into account my interpolation and the results of
Refs.~\cite{JLQCD_BB00,JLQCD_BBperturb}, as well as my estimate
of the systematic errors in both, I quote
$\hat B_{B_d,quench}=1.30(12)$.  It is not clear
what quenching error should be assigned here.  From 
quenched chiral perturbation theory (QChPT), Sharpe \cite{SHARPE_ICHEP98}
estimates a $10\%$ error.  There are not many simulations
that address this question.
Gimenez and Reyes \cite{GIMENEZandREYES_BS} find
a $\sim\!11\%$ difference between quenched
and unquenched results in the static-light case.  However, the systematic
effects in the two results are rather different, so it is difficult
to draw a strong conclusion.  MILC also has some
quite preliminary results \cite{MILC_BBprelim} in the static-light case,
suggesting a smaller quenching error, but again the systematics
in the comparison are not under good control.  At this point therefore
I quote the QChPT $10\%$ as the quenching error on the 
``world average'' \hbbd:
\begin{equation}
\ehbbd\! = \!1.30(12)(13);\ f_B\sqrt{\hat B_{B_d}}= \!230(40)\;\MeV,
\end{equation}
where the uncertainty on $f_B\sqrt{\hat B_{B_d}}$ is supposed to include
all sources of error.  
Since QChPT for \fb\ \cite{QChPT_fB} seems to 
overestimate somewhat the quenching errors, I
regard the assumed $10\%$  quenching effect on \hbbd\ as quite conservative.

For the ratio $\ehbbs/\ehbbd$, all groups get a number very close to 1.
Furthermore,  the experience with $\efbs/\efb$, as well as a
preliminary simulation \cite{MILC_BBprelim}, suggest that
the quenching errors on the ratio are considerably smaller
than the $\sim\!5\%$ inferred from QChPT \cite{SHARPE_ICHEP98}.
My world averages are then:
\begin{equation}
{\ehbbs\over \ehbbd} = 1.00(4);\ \; \xi\equiv{f_{B_s}\sqrt{\hat B_{B_s}}\over
f_{B_d}\sqrt{\hat B_{B_d}}}=1.16(5),
\label{eq:xiresults}
\end{equation}
where I have used Eq.~(\ref{eq:fBresults}).  All sources of error,
including quenching, are supposed to be represented in
Eq.~(\ref{eq:xiresults}).

Before concluding this section, I wish to mention results for 
$B_S$, the  B parameter for the scalar-pseudoscalar
$\Delta B=2$ operator, which is relevant for the Standard Model prediction
of the width difference $\Delta \Gamma/\Gamma$ for 
$B_s$ mesons.  $B_S$ has been computed with heavy quarks that 
are relativistic \cite{GUPTA_BS,APE_BS,APE_fB_BB00,BECIREVIC00},
nonrelativistic \cite{JLQCD_BB00,JLQCD_BBperturb}
and static \cite{GIMENEZandREYES_BS}.   The last reference includes 
a first look at unquenching effects.  The results from all
the groups appear consistent with $B^{\overline{MS}}_S(m_b)=0.82(8)$,
where I have defined  $B^{\overline{MS}}_S$ as in \cite{APE_BS}, and have
tried to include all sources of error.  
There are however large $1/m_Q$ corrections as well as disagreements
in how to go from $B_S$ to the width difference, so more
work is required before we have a Standard Model 
prediction for $\Delta \Gamma_{B_s}/\Gamma_{B_s}$.

\section{SEMILEPTONIC FORM FACTORS}
\label{sec:formfactor}

New work this year on semileptonic decays has dealt almost
exclusively with $B\to\pi\ell\nu$, so I will focus on that
process only.
The hadronic matrix element 
can be parameterized with form factors $f_+$ and $f_0$
defined by
\begin{eqnarray}
\quad\!\!\!\langle\pi(\vec k)|V_\mu|B(\vec p)\rangle &=&
f_+(q^2)\Big[p_\mu + k_\mu - \cr
\quad{(M_B^2-M_\pi^2)q_\mu\over q^2}\Big]\!\!\!\!\!
&+&\!\!\!\!f_0(q^2){(M_B^2-M_\pi^2)q_\mu\over q^2},
\end{eqnarray}
where $V_\mu$ is the vector current and $q\equiv p-k$ is the
4-momentum transfer to the leptons.

\begin{figure}[tb]
\null
\vspace{-.35truein}
\includegraphics[bb = 100  200 4096 4196,
width=2.7truein,height=2.3truein]{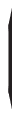}
\vskip -.1truein
\caption{}{Recent data for $B\to\pi\ell\nu$ formfactors $f_+$ and $f_0$.  
Only statistical errors are shown. }
\label{fig:formfactor}
\vskip -.1truein
\end{figure}

Figure~\ref{fig:formfactor} shows recent results for the form factors.
Various approaches have been tried:\footnote{A form factor study using
coarse anisotropic lattices \cite{SHIGEMITSU00} is not 
far  enough along yet to be included in this comparison.}
UKQCD \cite{UKQCD_BtoPI,MAYNARD}
and APE \cite{BECIREVIC00} use relativistic quarks and extrapolate to the
the B; JLQCD \cite{JLQCD_BtoPI} treats the heavy quark with NRQCD; and
the Fermilab group \cite{FERMILAB_BtoPI} uses their formalism
to simulate with the heavy quark at the $b$ mass.

In all the approaches, the first step in analysis of the raw
lattice data for the form 
factors (or for $\langle\pi(\vec k)|V_\mu|B(\vec p)\rangle$
in the Fermilab case) is an interpolation or fit. For APE or UKQCD,
this is a fit of $f_+$ and $f_0$ to expected phenomenological forms.
APE uses the nice ``B-K'' form \cite{B-K}, which automatically
incorporates the known scaling behavior in $1/M_{HL}$
at $q^2=q^2_{max}$ (the end point) and at $q^2=0$,
and enforces the kinematic constraint $f_+(0)=f_0(0)$. UKQCD employs
more conventional forms such as pole/dipole, although fits
to the B-K expressions do appear later in their analysis.
In practice, what is important at this stage is that the fit smooths out
data which is inherently rather noisy.  For example, I refer
the reader to Figure 2 in Ref.~\cite{MAYNARD}, which I believe
is typical of the kind of data seen by all the groups.  One must thus
keep in mind, when looking at plots like Fig.~\ref{fig:formfactor},
that the apparent smoothness of the data is a result of a procedure
that starts with an ``averaging'' process over a range in $q^2$ and
therefore produces highly correlated points.

The next step is a chiral extrapolation. In the heavy meson rest
frame, $q^2 = M_{HL}^2 + M_\pi^2 -2M_{HL}\sqrt{M_\pi^2+\vec k^2}$.
This means that an extrapolation of form factors at fixed, small $\vec k$
should contain a term linear in $M_\pi$, through the implicit $q^2$
dependence, in addition to the usual powers of $M_\pi^2$
\cite{UKQCD_chiralextrap}.  Alternatively, one can perform the extrapolation
at fixed $q^2$ \cite{UKQCD_BtoPI} or fixed 
$v\cdot k$ \cite{BECIREVIC00,JLQCD_BtoPI}
($v$ is the heavy meson 4-velocity), or simply extrapolate the B-K parameters
(or similar ones) rather than the form factors themselves \cite{BECIREVIC00}.
Yet another choice is to avoid the small $\vec k$ region altogether
(the decay rate is after all small there) and make a 
standard chiral extrapolation at fixed $\vec k$ \cite{FERMILAB_BtoPI}.

UKQCD and APE then need to extrapolate in $1/M_{HL}$ to the B mesons.
UKQCD extrapolates $f_+$ and $f_0$ at fixed $v\cdot k$, assuming that
$q^2$ is close enough to $q^2_{max}$ that the HQET forms valid near the
end point may be used.  APE extrapolates the B-K parameters themselves (and
also repeats the UKQCD-type analysis as a check).  The two
extrapolations are consistent within errors (see Fig.~\ref{fig:formfactor}),
suggesting that the HQET end-point forms are indeed appropriate.
Of course, just as for \fb\ and $B_B$, either extrapolation introduces
two sources of systematic error: (1) uncertainty in what order polynomial
(in $1/M_{HL}$) one should use, and (2) magnification of 
small $\cO(a^2M_{HL}^2)$
errors.  Both groups do an adequate job of estimating (1).  However, I think
that error (2) could use more study along the lines of 
Sec.\ \ref{sec:MILCvsUKQCDnorm}.

JLQCD and the Fermilab group work in the region of the B mesons,
so extrapolation in the heavy mass is not required.  Although
their results in Fig.~\ref{fig:formfactor} are similar, 
there may be some quantitative disagreement, especially in the
normalization of $f_0$ and in the $q^2$ dependences of both
form factors. I have some
concern that the Fermilab data for $f_+$ does not appear
to show the expected pole-like rise toward the end point seen
by the other groups. (Remember that each group's points are
highly correlated, so that the shape of the $q^2$ dependence is
probably highly significant.)  Of course, for APE and UKQCD, this
behavior is put in {\it ab initio} by the fit to the phenomenological
form.    JLQCD, on the other hand, does not appear to be forcing
this behavior by their analysis procedure, so the fact that it does
come out seems encouraging.

Since the Fermilab and JLQCD (NRQCD)
approaches  are fundamentally quite similar in the B meson region, it
is not easy to guess the source of the differences.  One
possibility is discretization error.  The Fermilab group (alone among the four) extrapolates
to the continuum from a range of $a$ ($\beta=5.7$, $5.9$, $6.1$),
while JLQCD works only at $\beta=5.9$.  However, the $a$-dependence
seen by the Fermilab group is mild, and it seems unlikely to me
that this is the explanation.

Perhaps a more likely culprit is the current renormalization.
JLQCD is working to one-loop order in perturbation theory. 
The Fermilab group does much of the renormalization nonperturbatively
(by comparing to the known normalization of the diagonal vector current),
but the remainder is still put in at tadpole-improved tree level for
the data in Fig.~\ref{fig:formfactor}.  They will be including 
the complete one-loop calculation shortly, and they assure me
that it makes little numerical difference.  It will, however, be interesting
to investigate the size of the renormalon shadows for both the
JLQCD and the Fermilab computations.

Another issue that has bedeviled lattice $B\to\pi\ell\nu$ calculations
for several years is the soft-pion relation  (SPR)
at the end point, $f_0(q^2_{max}) = f_B/f_\pi$ 
\cite{SOFTPI,BURDMAN}.  Some early calculations found
violations by as much as a factor of 2 \cite{EARLY_SOFTPI_PROB},
and although the situation has improved somewhat, the problem
has not gone away. 

At first sight,
it seems surprising that the end-point region should give trouble.
After all, lattice calculations are easiest at the end point where
no 3-momentum insertion is required ($\vec k= 0$ in the B rest frame).
However, the SPR is only guaranteed to be valid in the
limit $M_\pi\to0$.  As discussed above, the extrapolation in $M_\pi$
when $\vec k$ is fixed and small is tricky because $q^2$ changes and has a
term linear in $M_\pi \propto \sqrt{m_{quark}}$.  APE and UKQCD get
consistency with the SPR using chiral extrapolations
that avoid sitting at the end point (instead fixing $q^2$ 
or $v\cdot k$). The
end point is then reapproached after extrapolation to the 
physical $q^2_{max}$
and the B meson mass.  But, after all the extrapolations, 
the systematic errors are rather large, so the agreement with the SPR is
not very convincing.  Indeed, in the case of APE, the SPR is 
used as a criterion
to choose whether one should do linear or quadratic fits in $1/M_{HL}$,
so the agreement cannot then be seen as a success of the computation.  

\begin{figure}[tb]
\vspace{4pt}
\includegraphics[bb = 25  0 550 525,width=2.8truein,height=2.5truein]{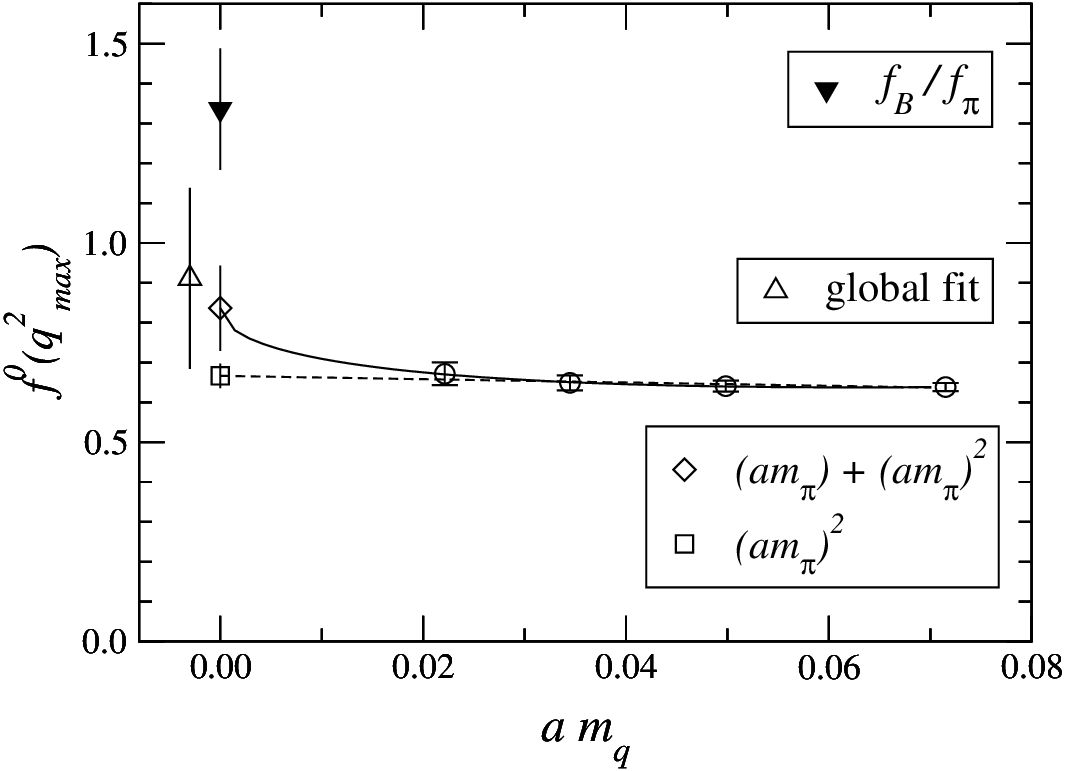}
\vspace{-1truein}
\caption{}{Extrapolation of $f_0$ at the end point to the chiral limit by
JLQCD \cite{JLQCD_BtoPI}.  The open triangle comes from a global fit to
the form in Ref.~\cite{BURDMAN}. 
}
\label{fig:softpi}
\end{figure}

JLQCD, on the other hand, studies the SPR in detail by sitting
at the end point with high statistics (2150 configurations).
Figure~\ref{fig:softpi} shows the chiral extrapolation of $f_0(q^2_{max})$.
Note that the fit (solid line) which takes into account the implicit linear
dependence on $M_\pi$ 
comes closer to $f_B/f_\pi$ than the purely quadratic fit in $M_\pi$ 
but still misses by a wide margin.  This illustrates how hard it will be
to verify the soft-pion relation directly:  the linear dependence
on $M_\pi$ causes a sharp rise, but only at very small
mass where there is no data.  This behavior is expected from the
effects of $q^2$ poles in the form factor in the unphysical
region just beyond the end point.  A global fit to the functional
dependence on $v\cdot k$ and $M_\pi$ expected from the
HQET-suggested form factors of Burdman \et\ \cite{BURDMAN},
does somewhat better, but still does not give convincing agreement
with $f_B/f_\pi$.    The Fermilab group has
independently decided to use the form factors of Ref.~\cite{BURDMAN}
to study the end-point behavior, and it will be very interesting
to see how their results compare to those of JLQCD.

Of course, one can always take the attitude that the SPR is
irrelevant to phenomenology because there is no rate near the end
point and, besides, it is $f_+$, not $f_0$, that determines
the rate into $\pi\mu\nu$ or $\pi e\nu$.  However, a 
credible demonstration of the
SPR would be very helpful in convincing the non-lattice community
that we have the form factor computation under control.

\section{SPECTRA}
\label{sec:spectra}

Although I do not have space for a detailed review of recent spectral
calculations for hadrons with one or more heavy quarks 
(Refs.~\cite{ALIKHAN_Qq,DRUMMOND_H,WOLOSHYN_Qq,HEIN_Qq,LEWIS_Qq,DAVIES_QQ,CPPACS_DYN_QQ,STEWART_QQ,CHEN_QQ,CPPACS_ANISO_QQ,MANKE_QQ_H,KRONFELD_HQET,KRONFELD_SIMONE}),
I will make some remarks about a few
features I find interesting and/or confusing:

(1) It seems clear that the computed quenched hyperfine splittings for 
heavy-light
mesons are considerably smaller than the experimental values.
For example, $M_{B_s^*}-M_{B_s}$ in the quenched
approximation appears to be about half its physical value
\cite{ALIKHAN_Qq,HEIN_Qq,LEWIS_Qq}, although some older calculations
\eg Ref~\cite{LEWIS_OLD},
found smaller ($\sim\!20\%$) discrepancies. For the
$D_s$  system, the effect is also present at about the $20\%$
level in most \cite{HEIN_Qq,DS_OLD,LEWIS_OLD} 
calculations (with one recent exception\cite{WOLOSHYN_Qq}). 
The sign and, very roughly, the
magnitude of the observed
discrepancy is expected from arguments about the effects of
quark loops on the
short distance potential.
However, I know of no direct evidence from simulations that
quenching is the culprit.

(2) There is a disagreement between Refs.~\cite{LEWIS_Qq} and \cite{ALIKHAN_Qq}
in the computed P-wave splittings
for B mesons ($M_{B^*_2} - M_{B^*_0}$ or  $M_{B^*_0} - M_{B}$,
 or the corresponding quantities
for $B_s$).  The disagreement is considerably larger than the
systematic errors quoted by both groups and deserves further
study.    My guess is that the problem lies in excited state
contamination:  These mass differences are quite noisy, and it is
seems very difficult to identify a ``true plateau'' in
the effective mass plots.

(3) For NRQCD computations of splittings in charmonium,
Stewart and Koniuk \cite{STEWART_QQ} confirm previous evidence \cite{TROTTIER}
that the results are very sensitive to the order (in the velocity $v$)
to which the Hamiltonian is corrected and how the tadpole improvement
is done (plaquette or Landau link).   Indeed, including the $\cO(v^6)$
terms, with the renormalization done at tadpole-improved tree level,
moves the computed hyperfine splitting significantly {\it further} from
experiment.  I believe this is a ``smoking gun'' for renormalon shadow
effects: At tadpole-improved tree level the presence of higher
dimensional operators has no effect on the coefficients of lower
dimensional operators --- there is no subtraction of power law divergences.
Just as for $f_B$, putting in the higher dimensional operators
without the subtraction is then worse than leaving out those
operators entirely. Unlike the \fb\ case, however, the effect here
appears to be numerically dramatic.  

(4) In the $b\bar b$ system, NRQCD is clearly better behaved than
in $c\bar c$, since the relativistic corrections are smaller.
It is therefore possible to find convincing evidence
for quenching effects \cite{CPPACS_ANISO_QQ,DAVIES_QQ} if one 
is careful to keep other systematic effects
(lattice spacing, order in $v$, type of tadpole improvement) fixed
when comparing quenched and unquenched simulations.  However, even
for $b\bar b$, I believe it is important to reduce renormalon shadows
by doing at least one-loop renormalization of the Hamiltonian
before one can make reliable comparisons with experiment for most
quantities.

(5) Following pioneering work by Klassen \cite{KLASSEN}, there 
have been several recent
simulations of charmonium on quenched ``anisotropic relativistic''
lattices \cite{CHEN_QQ,CPPACS_ANISO_QQ,MANKE_QQ_H}. The anisotropy
$\xi$ is defined by $\xi\equiv a_s/a_t$, where $a_s$ and $a_t$ are
the spatial and temporal lattice spacings, respectively.
Choosing $\xi>1$ (typically $2\le \xi \le 5$)
allows one to keep $a_tm_Q\ll 1$  and 
avoid large mass dependence in the improvement coefficients.
Effectively, the heavy quark is 
relativistic.\footnote{
To see that $a_s M\!\sim \!1$
does not introduce uncontrolled mass dependence requires \cite{KRONFELD_REMARK} 
an analysis along the
lines of \cite{EKM}.}
Although the improvement coefficients through $\cO(a)$
can in principle be evaluated nonperturbatively, in practice
the spatial and temporal clover coefficients are determined
at tadpole-improved tree level, either by the Landau link or the
plaquette.  

\begin{figure}[tb]
\includegraphics[width=7.4truecm,height=2.1truein]{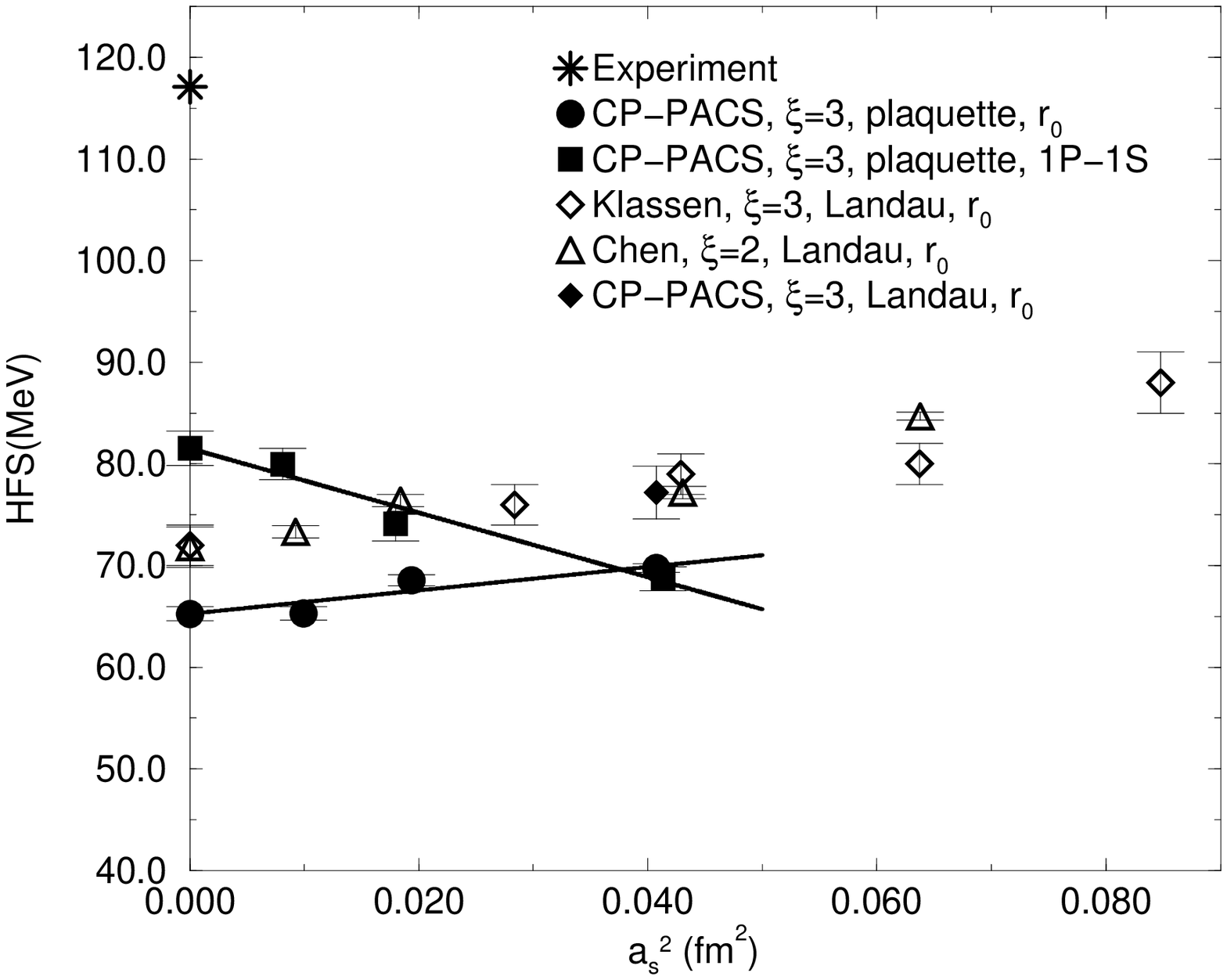}
\vskip -.3truein
\caption{}{Hyperfine splitting in charmonium {\it vs.}\ $a_s^2$ for 
relativistic anisotropic lattices.
Data is from
Refs.~\cite{KLASSEN,CHEN_QQ,CPPACS_ANISO_QQ}; figure is taken from
\cite{CPPACS_ANISO_QQ}.  
}
\label{fig:aniso_hfs}
\vskip -.15truein
\end{figure}

Figure~\ref{fig:aniso_hfs} shows the continuum
extrapolation for the charmonium hyperfine splitting for various
choices of anisotropy, tadpole factor, and scale determination.
The good news is that the results from this method seem
quite precise compared
to previous relativistic calculations on isotropic lattices \cite{PREVIOUS_HFS}
and do not have the sensitive dependence on the tadpole factor seen 
\cite{STEWART_QQ,TROTTIER} in NRQCD.  The bad news is that there
appears to be some remaining dependence on the tadpole factor in the continuum
limit (compare the open triangles and filled circles).\footnote{
Differences in the continuum value due
to different scale choices ($r_0$ or the 1P-1S splitting) are presumably
explained by quenching and/or phenomenological uncertainties in
$r_0$ (see Sec.~\ref{sec:fB}).}
The reason for the discrepancy is
not yet clear.  Additional data at larger $a_s$ with the plaquette tadpole 
may be useful in sorting out what is happening.

(6) The anisotropic relativistic  approach has also been quite
helpful in getting good signals for quarkonium hybrids \cite{MANKE_QQ_H}.
As in glueball calculations \cite{MORNINGSTAR}, anisotropy gives
more usable time slices before the 
signal is lost in noise.  Another recent hybrid calculation \cite{DRUMMOND_H}
also employs anisotropic lattices, but in the NRQCD context.

(7) Kronfeld \cite{KRONFELD_HQET} has recently argued that a heavy
quark on the lattice can be represented by a {\it continuum} HQET
where the dependence on $am_Q$ is completely absorbed into the
short-distance coefficients of the heavy quark operators.
The operator matrix elements then only have mild lattice spacing
dependence though the light degrees of freedom. This nice insight
is then applied \cite{KRONFELD_SIMONE} to the heavy-light meson spectrum 
on the lattice in an attempt to compute the HQET parameters
$\bar\Lambda$ and $\lambda_1$. ($\bar\Lambda$ is the meson
binding energy in the static limit; $-\lambda_1/2m_Q$, the heavy quark
kinetic energy.)  

I have some concerns about Ref.~\cite{KRONFELD_SIMONE}.
One could find the continuum HQET by a two-step
process: (a) for fixed $a$ in the lattice theory, take $m_Q$ large
and arrive at a {\it lattice} HQET, and (b) order by order in perturbation
theory, replace the lattice regularized HQET operators by their continuum
counterparts.
Although this not the procedure employed in
\cite{KRONFELD_SIMONE}, I believe the approaches should be equivalent.
If I am right, then there are effective power law divergences
and renormalon
shadow effects\footnote{
In the Fermilab formalism, there are no true
power law divergences (aside from the standard additive shift in 
the mass) as $a\to0$ for fixed $m_Q$ because the relativistic theory
is regained. The coefficients (\eg $d_1$) vanish fast enough
to cancel the power divergence in the higher dimensional operators.
However, as discussed for $f_B$ in Sec.~\ref{sec:renormalon},
the power divergences of the operators themselves imply that
the renormalon shadow issue is still present.}
in the coefficients of the continuum operators
(resulting from step (b)). Therefore, a one-loop calculation
may not be sufficiently accurate to extract $\bar\Lambda$ and $\lambda_1$.
Indeed, parameterizing the discretization errors on $\bar\Lambda$ 
as $\bar \Lambda(a)=\bar \Lambda(0)(1-a\cM)$, 
the slope $\cM$ has a value of $\approx\!0.65\;\GeV$.
Yet, if the lattice spacing dependence is coming only from the light degrees
of freedom, which are at least tree-level improved through $\cO(a)$,
one expects $\cM = \cO(\alpha_s \Lambda_{QCD})$, an order of magnitude
smaller than what is found. This may be a renormalon shadow effect.

\section{CONCLUDING REMARK}
\label{sec:conclusions}

\begin{figure}[tb]
\includegraphics[width=3.1truein,height=1.8truein]{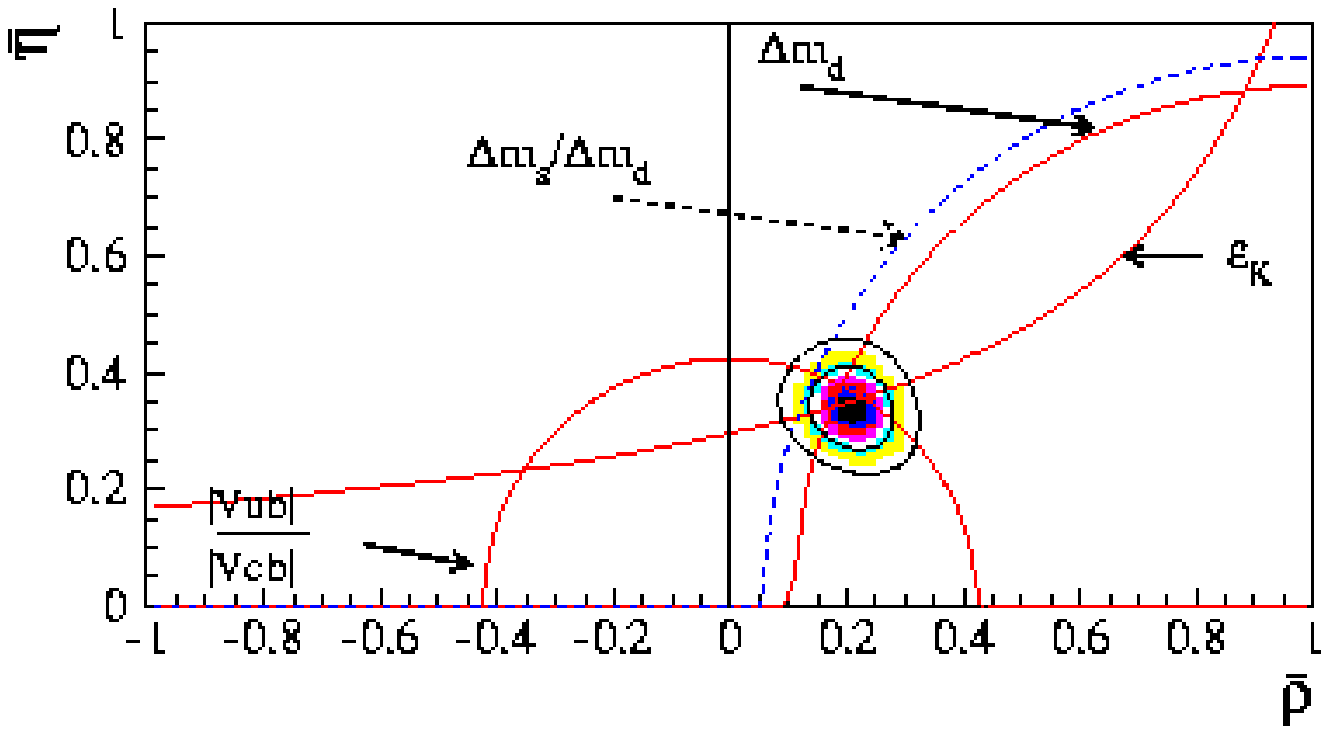}
\vskip -.3truein
\caption{}{Allowed region for the vertex of the unitary triangle 
(CKM parameters $\bar \rho$ and $\bar \eta$)
from Ref.~\cite{TRIANGLE00}.}
\label{fig:triangle}
\vskip -.15truein
\end{figure}

I have not made a unitarity triangle analysis with my world average
lattice results.  However, Fig.~\ref{fig:triangle} shows the result
of an analysis \cite{TRIANGLE00} that employs lattice values 
similar to those quoted here.\footnote{The main difference is that \cite{TRIANGLE00}
uses the value
$f_B\sqrt{\hat B_{B_d}}= 220(25)(20)\;\MeV$ rather than my 
$230(40)\;\MeV$.}   Clearly, our knowledge
of the triangle is getting quite precise.
Over the last dozen years, the allowed
region for the vertex has shrunk in area by almost a factor of 20,
as is shown dramatically in
Fig.~4 of Ref.~\cite{TRIANGLE_EVOLUTION}.
We can be proud of the fact that
lattice computations have played an important role
in this accomplishment.

\section{ACKNOWLEDGEMENTS}

I thank
A.\ Ali Khan
G.\ Bali,
D.\ Becirevic,
R.\ Burkhalter,
S.\ Gottlieb,
R.\ Gupta,
J.\ Hein,
K.-I.\ Ishikawa,
A.\ Kronfeld,
L.\ Lellouch,
R.\ Lewis,
P.\ Mackenzie,
T.\ Manke,
G.\ Martinelli,
C.\ Maynard,
C.\ McNeile,
M.\ Okamoto,
T.\ Onogi,
S.\ Ryan,
H.\ Shanahan,
S.\ Sint,
R.\ Sommer
and N.\ Yamada
for discussion and private communication.
This work was supported in part
by the US Department of Energy under grant DE-FG02-91ER40628.


\begin{thebibliography}{9}
\bibitem{LUBICZ} V.\ Lubicz, these proceedings.

\bibitem{BALI} G.\ Bali, hep-ph/0001312.

\bibitem{HASHIMOTO_LAT99} S.\ Hashimoto, \pisa\ 3.

\bibitem{DRAPER_LAT98} T.\ Draper, \boulder\  43.

\bibitem{FNAL97} A.\ El-Khadra \et, \prd{58} (1998) 014506.

\bibitem{APE97} C.\ Allton \et, \plb{405} (1997) 133.

\bibitem{JLQCD98} S.\ Aoki \et, \prl{80} (1998) 5711.

\bibitem{MILC_PRL} C.\ Bernard \et,
\prl{81} (1998) 4812.

\bibitem{ALIKHAN98} A.\ Ali Khan \et, \plb{427} (1998) 132.

\bibitem{JLQCD99}  K-I.\ Ishikawa \et, \prd{61} (2000) 074501.

\bibitem{APE99} D.\ Becirevic \et, \prd{60} (1999) 074501.

\bibitem{APE_fB_BB00} D.\ Becirevic \et, hep-lat/0002025.

\bibitem{UKQCD_fB00} K.C. Bowler, \et, hep-lat/0007020.

\bibitem{MAYNARD} C.\ Maynard for UKQCD, these proceedings (hep-lat/0010016) and private
communication.

\bibitem{SOMMER93} R.\ Sommer, \npb{411} (1994) 839. 
\bibitem{MILC_LAT00} S.\ Datta for the MILC collaboration,
these proceedings (hep-lat/0011029).

\bibitem{CPPACS_fB00} A.\ Ali Khan for CP-PACS, these proceedings;
A.\ Ali Khan and H.\ Shanahan, private communications.

\bibitem{CPPACS_NEW} A.\ Ali Khan \et, hep-lat/0010009.

\bibitem{LELLOUCH00} L.\ Lellouch and C.-J.\ Lin,
private communication and hep-ph/0011086; 
contribution to {\it Heavy Flavors 8}, Southampton,
England, 25-29 Jul 1999, hep-ph/9912322.

\bibitem{COLLINS99} S.\ Collins \et, \prd{60} (1999) 074504.



\bibitem{BECIREVIC00} 
D.\ Becirevic for APE, these proceedings and private communication;
\pisa\ 268.

\bibitem{IWASAKI} Y.\ Iwasaki, \npb{258} (1985) 141.

\bibitem{BURKHALTER} R.\ Burkhalter for CP-PACS, these proceedings (hep-lat/0010078)
and private communication. 

\bibitem{EICHTEN87} E.\ Eichten, \seillac\ 170.

\bibitem{LEPAGE87} G.P.\ Lepage and B.A.\ Thacker, \seillac\ 199.

\bibitem{EKM} A.\ El-Khadra, A.\ Kronfeld and
P.\ Mackenzie, \prd{55} (1997) 3933.

\bibitem{MARTINELLIandSACHRAJDA96} G.\ Martinelli and C.\ Sachrajda,
\npb{478} (1996) 660. 

\bibitem{OTHERRENORMALON} M.\ Beneke, Phys.\ Rept.\ 317 
(1999) 1; 
G.\ Bodwin and Y.-Q.\ Chen, \prd{60} (1999) 054008. 

\bibitem{MARTINELLIandSACHRAJDA98} G.\ Martinelli and C.\ Sachrajda,
\npb{559} (1999) 429. 

\bibitem{LEPMAC}  G.P.\ Lepage and P.\ Mackenzie, \prd{48} (1993) 2250.

\bibitem{ALPHA_CSW-CA-ZA} M.\ L\"uscher, \et, \npb{491} (1997) 323;
{\it ibid}, 344.


\bibitem{IOY} K.-I.\ Ishikawa, T.\ Onogi, and N.\ Yamada,
\pisa\ 301. 

\bibitem{ISHIKAWA} I thank K.-I.\ Ishikawa for providing me with
their results for this subtraction coefficient.

\bibitem{CB-TD} C.\ Bernard and T.\ DeGrand, in preparation.

\bibitem{BOOSTEDPT} G.\ Parisi, in {\it High Energy Physics --- 1980},
L.\ Durand and L.G.\ Pondrum, eds.,  (AIP, New York, 1981).

\bibitem{COLLINS00} S.\ Collins \et, hep-lat/0007016.

\bibitem{HandH} O.\ Hernandez and B.\ Hill,
\prd{50} (1994) 495. 

\bibitem{KURAMASHI}
Y.\ Kuramashi,
\prd{58} (1998) 034507.

\bibitem{BHATTACHARYA99} T.\ Bhattacharya \et, \pisa\ 851.

\bibitem{SINTandWEISZ} S.\ Sint and P.\ Weisz, \npb{502} (1997) 251.

\bibitem{BHATTACHARYA00} T.\ Bhattacharya \et, 
hep-lat/0009038; talks by T.\ Bhattacharya and R.\ Gupta, these proceedings.

\bibitem{THANKSINT}  I thank S.\ Sint for pointing out an incorrect use
of the equations of motion in an
earlier version of NP-tad.

\bibitem{SOMMER00} M.\ Kurth and R.\ Sommer, hep-lat/0007002;
J.\ Heitger, M.\ Kurth and R.\ Sommer, in preparation and
private communication.

\bibitem{MILC_LAT99} C.\ Bernard \et, \pisa\ 289.

\bibitem{APESMEAR}
M.\ Albanese \et,
\plb{192} (1987) 163.

\bibitem{CB-TD_LAT99} C.\ Bernard and T. DeGrand, \pisa\ 845.

\bibitem{STEPHENSON} M.\ Stephenson \et, hep-lat/9910023.

\bibitem{GIMENEZandREYES} V.\ Gimenez and J.\ Reyes,
\npb{545} (1999) 576 
[raw data is taken from UKQCD (A.K.\ Ewing \et),
\prd{54} (1996) 3526 and APE (V.\ Gimenez and G.\ Martinelli) \plb{398} (1997) 135].

\bibitem{CDM} J.\ Christensen, T.\ Draper, and C.\ McNeile,
\prd{56} (1997) 6993.

\bibitem{JLQCD_BB00} N.\ Yamada for JLQCD, these proceedings (hep-lat/0010089)
and private
communication.

\bibitem{JLQCD_BBperturb} K.-I.\ Ishikawa \et, hep-lat/0004022
and these proceedings (hep-lat/0010056).

\bibitem{JLQCD_BB99} S.\ Hashimoto \et, \prd{60} (1999) 094503;
\prd{62} (2000) 034504.


\bibitem{SHARPE_ICHEP98} S.\ Sharpe, talk at ICHEP 98, Vancouver, Canada,
July, 1998,  hep-lat/9811006.

\bibitem{GIMENEZandREYES_BS} V.\ Gimenez and J.\ Reyes, these
proceedings (hep-lat/0010048); hep-lat/0009007.

\bibitem{MILC_BBprelim} MILC collaboration, work in progress.

\bibitem{QChPT_fB} M.\ Booth, \prd{51} (1995) 2338; S.\ Sharpe
and Y.\ Zhang, \prd{53} (1996) 5125.

\bibitem{GUPTA_BS} R.\ Gupta \et, \prd{55} (1997) 4036.

\bibitem{APE_BS} D.\ Becirevic \et, hep-ph/0006135.

\bibitem{UKQCD_BtoPI} K.C.\ Bowler \et, \plb{486} (2000) 111;
C.\ Maynard for UKQCD, \pisa\ 322.

\bibitem{JLQCD_BtoPI} T.\ Onogi for JLQCD, these proceedings (hep-lat/0011008)
and private communication.

\bibitem{FERMILAB_BtoPI} S.\ Ryan,  A.\ Kronfeld and P.\ Mackenzie,
private communications; S.\ Ryan \et, \pisa\ 328.

\bibitem{SHIGEMITSU00} J.\ Shigemitsu \et, these proceedings (hep-lat/0010029).

\bibitem{B-K} D.\ Becirevic and A.\ Kaidalov,
\plb{478} (2000) 417.

\bibitem{UKQCD_chiralextrap}  K.C.\ Bowler \et, \prd{51} (1995) 4905.

\bibitem {SOFTPI}   G.\ Burdman and J.F.\ Donoghue, \plb{280}
(1992) 287; M.B.\ Wise, \prd{45} (1995) 2188; N.\ Kitazawa and T.\ Kurimoto
\plb{323} (1994) 65. 

\bibitem {BURDMAN} G.\ Burdman \et, \prd{49} (1994) 2331.

\bibitem{EARLY_SOFTPI_PROB} H.\ Matsufuru \et, \edinburgh\ 368;
S.\ Aoki, {\it ibid}, 380.

\bibitem{ALIKHAN_Qq} A.\ Ali Khan \et, \prd{62} (2000) 054505.

\bibitem{DRUMMOND_H} I.T.\ Drummond \et, \plb{478} (2000) 151.

\bibitem{WOLOSHYN_Qq} R.M.\ Woloshyn, \plb{476} (2000) 309.

\bibitem{HEIN_Qq} J.\ Hein \et, \prd{62} (2000) 074503 
and private communication.

\bibitem{LEWIS_Qq} R.\ Lewis and R.M.\ Woloshyn, hep-lat/0003011; these
proceedings (hep-lat/0010001) and private communication.

\bibitem{DAVIES_QQ} C.\ Davies, these proceedings (L.\ Marcantonio
\et, hep-lat/0011053). 

\bibitem{CPPACS_DYN_QQ}  T.\ Manke \et, hep-lat/0005022.

\bibitem{STEWART_QQ} C.\ Stewart and R.\ Koniuk,  hep-lat/0005024
and these proceedings (hep-lat/0010015).

\bibitem{CHEN_QQ} P.\ Chen, hep-lat/0006019.

\bibitem{CPPACS_ANISO_QQ}  M.\ Okamoto for CP-PACS, these proceedings
(hep-lat/0011005) and private communication.

\bibitem{MANKE_QQ_H} P.\ Chen, X.\ Liao and T.\ Manke, these
proceedings (hep-lat/0010069); T.\ Manke, private communication.

\bibitem{KRONFELD_HQET} A.\ Kronfeld, \prd{62} (2000) 014505 and
private communication.

\bibitem{KRONFELD_SIMONE} A.\ Kronfeld and J.\ Simone,
\plb{490} (2000) 228.

\bibitem{LEWIS_OLD} R.\ Lewis and R.M.\ Woloshyn, \prd{58} (1998) 074506.

\bibitem{DS_OLD} P.\ Mackenzie, \edinburgh\ 305; P. 
Boyle (UKQCD), {\it ibid}, 314.

\bibitem{TROTTIER} H.D.\ Trottier, \prd{55} (1997) 6844;
N.H.\ Shakespeare and H.D.\ Trottier,
\prd{58} (1998) 034502.

\bibitem{KLASSEN} T.\ Klassen,
\npb{533} (1998) 557;
\boulder\  918;
and unpublished.


\bibitem{KRONFELD_REMARK} 
I thank A.\ Kronfeld for this remark.

\bibitem{PREVIOUS_HFS} Reference \cite{CHEN_QQ} compares, for example,
to unpublished work from the Fermilab group.

\bibitem{MORNINGSTAR} C.\ Morningstar and M.\ Peardon,
\prd{56} (1997) 4043.

\bibitem{TRIANGLE00} M.\ Ciuchini \et, paper submitted to 
ICHEP 2000, Osaka, 27 July -- 2 August, 2000. 

\bibitem{TRIANGLE_EVOLUTION} F.\ Caravaglios \et, hep-ph/0002171.

\end{thebibliography}
\end{document}